\documentclass[aoas, preprint,authoryear]{imsart}

\RequirePackage[OT1]{fontenc}
\RequirePackage{amsthm,amsmath, amssymb}
\RequirePackage[colorlinks,citecolor=blue,urlcolor=blue]{hyperref}
\RequirePackage[numbers]{natbib}

\usepackage{graphicx}
\usepackage{caption}
\usepackage{subcaption}
\graphicspath{ {./figures/} }

\usepackage{bm}
\usepackage{dsfont} 

\usepackage{lmodern,textcomp} 

\usepackage{pifont}
\newcommand{\cmark}{\ding{51}}%
\newcommand{\xmark}{\ding{55}}%

\usepackage{tabularx}

\usepackage{multirow}

\usepackage{makecell}

\usepackage{xr} 
\usepackage{array} 

\usepackage{tikz} 

\usetikzlibrary{shapes,decorations,arrows,calc,arrows.meta,fit,positioning}
\tikzset{
    -Latex,auto,node distance =1 cm and 1 cm,semithick,
    partial/.style ={regular polygon,regular polygon sides=4, draw, dashed, minimum width = 0.7 cm, inner sep=3pt},
    intervene/.style={regular polygon, regular polygon sides=3, rotate=30, draw, minimum size = 0.1 cm, inner sep=0.5pt},
    state/.style ={regular polygon,regular polygon sides=4, draw, minimum width = 0.7 cm, inner sep=3pt},
    missing/.style ={ellipse, draw, dashed,minimum width = 0.7 cm},
    point/.style = {circle, draw, inner sep=0.04cm,fill,node contents={}},
    dasheddirected/.style={Latex,dashed},
    el/.style = {inner sep=2pt, align=left, sloped}
}
\usetikzlibrary{bayesnet}
\usepackage{rotating}
\usepackage{stefan_tex}

\startlocaldefs

\DeclareMathOperator*{\logit}{logit}
\newcommand\independent{\protect\mathpalette{\protect\independenT}{\perp}} 
\def\independenT#1#2{\mathrel{\rlap{$#1#2$}\mkern2mu{#1#2}}}

\newcommand{\na}{\operatorname{NA}}

\endlocaldefs

\usepackage[colorinlistoftodos,textwidth=2.3cm]{todonotes}

\MakeRobust{\ref}
\makeatletter
\newcommand{\labeltext}[2]{%
  \@bsphack
  \csname phantomsection\endcsname 
  \def\@currentlabel{#1}{\label{#2}}%
  \@esphack
}
\makeatother

\begin{document}

\begin{frontmatter}

\title{Doubly robust treatment effect estimation with missing attributes}
\runtitle{ATE estimation with missing attributes}

\begin{aug}


\author{\fnms{Imke} \snm{Mayer}\corref{}\ead[label=e1]{imke.mayer@ehess.fr}},
\author{\fnms{Erik} \snm{Sverdrup}\ead[label=e2]{erikcs@stanford.edu}},
\author{\fnms{Tobias} \snm{Gauss}\ead[label=e3]{tgauss@protonmail.com}}, \\
\author{\fnms{Jean-Denis} \snm{Moyer}\ead[label=e4]{jean-denis.moyer@aphp.fr}},
\author{\fnms{Stefan} \snm{Wager}\ead[label=e5]{swager@stanford.edu}}
\and
\author{\fnms{Julie} \snm{Josse}\ead[label=e6]{julie.josse@polytechnique.edu}}

\affiliation{\'Ecole des Hautes \'Etudes en Sciences Sociales,
Stanford University,
Beaujon Hospital and
\'Ecole Polytechnique} 

\address{I.~Mayer \\
Centre d'Analyse et de Math\'ematique Sociales\\
\'Ecole des Hautes \'Etudes en Sciences Sociales \\
75006 Paris, France \\
\printead{e1}}

\address{E.~Sverdrup and S.~Wager \\
Graduate School of Business\\
Stanford University\\
CA 94305, USA\\
\printead{e2} \\
\printead{e5}}

\address{T.~Gauss and J.-D.~Moyer \\
Department of anesthesia and intensive care\\
Beaujon hospital, AP--HP\\
92110 Clichy, France\\
\printead{e3} \\
\printead{e4}}

\address{J.~Josse \\
Centre de Math\'ematiques Appliqu\'ees\\
\'Ecole Polytechnique\\
Institut Polytechnique de Paris\\
91128 Palaiseau Cedex, France\\
\printead{e6}}

\runauthor{I. Mayer et al.}

\end{aug}

\begin{abstract}

Missing attributes are ubiquitous in causal inference, as they are in most applied statistical work.
In this paper, we consider various sets of assumptions under which causal inference is possible
despite missing attributes and discuss corresponding approaches to average treatment effect estimation,
including generalized propensity score methods and multiple imputation.
Across an extensive simulation study, we show that no single method  systematically out-performs others.
We find, however, that doubly robust modifications of standard methods for average
treatment effect estimation with missing data repeatedly perform better than their non-doubly robust baselines;
for example, doubly robust generalized propensity score methods beat inverse-weighting with the generalized propensity score.
This finding is reinforced in an analysis of an observational study on the effect on mortality of tranexamic
acid administration among patients with traumatic brain injury in the context of critical care management.
Here, doubly robust estimators recover confidence intervals that are consistent with evidence from randomized
trials, whereas non-doubly robust estimators do not.

\end{abstract}

\begin{keyword}[class=MSC]
\kwd[Primary ]{93C41} 
\kwd{62G35} 
\kwd{62F35} 
\kwd[; secondary ]{62P10} 
\end{keyword}

\begin{keyword}
\kwd{causal inference}
\kwd{potential outcomes}
\kwd{observational data}
\kwd{propensity score estimation}
\kwd{incomplete confounders}
\kwd{major trauma}
\kwd{public health}
\end{keyword}

\end{frontmatter}

\section{Introduction}
\subsection{Hemorrhagic shock and traumatic brain injury in critical care management}
Our work is motivated by a prospective observational study of the causal effect of tranexamic acid (TA), an antifibrinolytic
agent that limits excessive bleeding, on mortality among traumatic brain injury patients
during their stay at the hospital (from admission to ICU and regular care units).
The beneficial effect of TA on mortality has been shown in a large randomized placebo-controlled study
\citep{shakur_CRASH2010}. Our interest in developing observational study methods for assessing the
effect of TA is twofold: In the long run, observational studies will be able to incorporate data on a larger and
more diverse set of patients, thus allowing us to get a better understanding of when and for whom TA
works; and treatment effect estimation on such observational studies can serve as a precursor 
for future randomized placebo-controlled studies, namely by helping defining the most interesting 
or promising target population beforehand and the associated inclusion rules.

Our study is built on top of the Traumabase\textsuperscript{\textregistered} database, which currently indexes
around 20,000 major trauma patients.\footnote{Major trauma is defined as any injury that potentially
causes prolonged disability or death and it is a public health challenge and a major source of mortality
and handicap around the world \citep{hay_etal_2017}.}
For each patient, 244 measurements are collected both before and during the hospital stay, including
both quantitative and categorical variables.
As shown in Table \ref{tab:start}, TA was administered to roughly 8\% of traumatic brain injury patients, and among all patients 20\%
died before the end of their hospital stay. We also see that mortality was much higher among patients who received
TA than those who did not (46\% vs. 18\%). This apparent reversal of the expected causal effect is a standard example of confounding bias
(also known as Simpson's paradox): The effect arises because patients who appeared to be in more severe state were more
likely to be administered TA and were also more likely to die with or without the treatment.

\begin{table}
\begin{tabular}{r|cc|}
     & survived & died \\ \hline
TA not administered & 6,238 (76\%) & 1,327 (16\%) \\ 
TA administered & 367 (4\%) & 316 (4\%) \\ \hline 
\end{tabular}
\caption{Occurrence and frequency table for traumatic brain injury patients (total number: 8,248).}
\label{tab:start}
\end{table}

The goal of our observational study design is to use a subset of 37 auxiliary covariates collected by the
Traumabase group to control for confounding and identify the causal
effect of TA on mortality. This ``unconfoundedness'' or ``selection on observables'' strategy is justified
if the treatment of interest (i.e., administration of TA) is as good as random after conditioning on covariates
\citep{imbens_rubin_2015,rosenbaum_rubin_1983}. In general, such an unconfoundedness assumption cannot be
validated from data, and needs to be built into the observational study design.

In order to make unconfoundedness
as plausible as possible, the Traumabase group chose which covariates among the total of 244 collected covariates to 
incorporate in our study
by soliciting feedback from a number experts using the Delphi method \citep{dalkey_helmer_1963, jones_hunter_1995}.
The focus of the Delphi survey was in understanding which factors were important for understanding health trajectories
of major trauma patients. Because the decision whether or not to administer TA was performed
by health professionals, it is likely that this same set of variables is also relevant to understanding which patients were
more likely than others to be selected for treatment.
A detailed list of the confounders and predictors of the outcome, in-ICU mortality, that were
chosen via the Delphi method is given in the \ref{suppA}. 

As discussed further in the following section, the statistics of treatment effect estimation under unconfoundedness
are by now well understood, with literature covering a range of topics from identification
\citep{imbens_rubin_2015,rosenbaum_rubin_1983} and simple weighted estimators
\citep{abadie2016matching,rosenbaum_rubin_JASA1984,zubizarreta2012using} to
semiparametrically efficient estimation in potentially high-dimensional settings
\citep{athey2018approximate,chernozhukov_etal_2018,robins_etal_JASA1994,van2011targeted}
and optimal treatment personalization \citep{athey2017efficient,kitagawa2018should,luedtke2016statistical,zhao2012estimating}.

\begin{figure}[t]
\begin{center}
\includegraphics[width=0.8\textwidth]{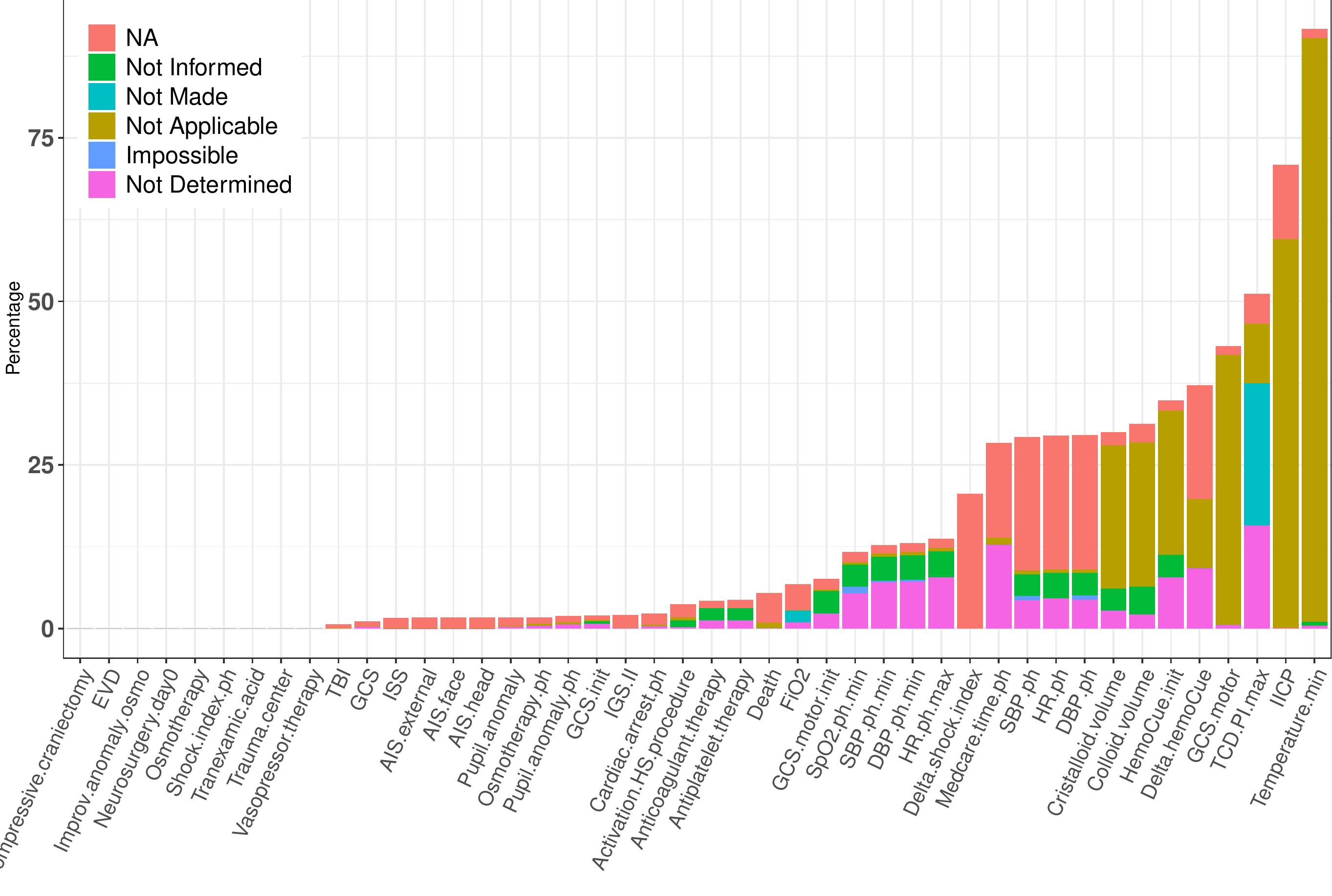}
\caption{Percentage of missing values for a subset of variables relevant for traumatic brain injury. Different encodings of missing values: \textit{NA} (not available), \textit{not informed}, \textit{not made}, \textit{not applicable}, \textit{impossible}.}
\label{fig:traumabase-missing}
\end{center}
\end{figure}

In the case of the Traumabase dataset, however, we have an additional complication whereby,
in Figure \ref{fig:traumabase-missing}, many of the variables have missing entries.
Some of the missingness is presumably due to non-informative missingness, e.g., medical
staff simply forgetting to log some numbers, but in other cases the missingness is clearly informative;
and in fact the analysts compiling the dataset used many different phrases to describe missing measurements,
ranging from ``not made'' and ``not applicable'' to ``impossible''. The last denomination arises, for example,
in the case of blood pressure measurements for patients in cardiac arrest or with dismemberment, as first responders
simply cannot measure blood pressure for patients suffering from one of these two conditions.
Meanwhile, variables indicating the response to a certain drug, such as the pupil contraction after the administration of
a saline solution, systematically take on the value ``not applicable'' if the treatment has not been administered
(the latter is informed in a separate variable). 

There are a handful of popular strategies for working with missing values in the context of treatment effect estimation under
unconfoundedness, ranging from generalized propensity score methods \citep{dagostino_rubin_JASA2000,rosenbaum_rubin_JASA1984}
to multiple imputation \citep{little_rubin_2002,rubin_1976, rubin_MINS1987}. However, the methodology for treatment
effect estimation with missingness is not as thoroughly fleshed out as corresponding methods without missing data.
In particular, although doubly robust and semiparametrically efficient methods have shown considerable promise in
cases without missingness \citep{athey2018approximate,chernozhukov_etal_2018,robins_etal_JASA1994,van2011targeted},
we are not aware of a study of doubly robust treatment effect methods with missing covariates.

\subsection{Summary of contributions and outline}

In this paper, we consider several popular methods for treatment effect estimation
with missing covariates that rely on various unconfoundedness assumptions or assumptions about the 
missingness mechanism. We then discuss natural doubly robust generalizations of these methods,
and compare them in numerical experiments. We find considerable variability in which methods
perform best in our experiments. Sometimes methods that start from generalized propensity scores do better, while other times
multiple imputation with parametric methods fit via the EM algorithm \citep{dempster_etal_1977} are better whereas other times non-parametric
estimators do better; overall, the performance of each method strongly depends on the underlying confounding mechanism.
However, we systematically find our doubly robust modifications of standard methods to outperform their baselines.

In the case of the Traumabase study, all doubly robust estimators give confidence intervals that cover 0, indicating
that we need to collect more data before we can use the observational study to guide clinical choices around administration of TA in the context of traumatic brain injury.
In contrast, all baseline methods result in confidence intervals that do not cover 0, and find significantly harmful effects of TA
on mortality.
It thus appears that using doubly robust estimators is needed to eliminate the
selection bias seen in Table \ref{tab:start}.

\section{Methods for Complete Data}

As a preliminary to our discussion on how to estimate causal effects with missing attributes,
we first briefly review methods that are widely used in the easier case without missingness.
Suppose we observe $n$ independent and identically distributed samples
$(X_i, \, Y_i, \, W_i) \in \RR^p \times \RR \times \cb{0, \, 1}$  where $X_i$ is a vector
of attributes, $Y_i$ is an outcome of interest, and $W_i$ denotes treatment assignment.
We define causal effects via the Neyman-Rubin potential outcomes model
under the stable unit treatment value assumption \citep{imbens_rubin_2015}.
We posit potential outcomes $\cb{Y_i(0), \, Y_i(1)}$ corresponding to the outcome the $i$-th sample would have experienced
had they been assigned treatment $W_i = 0$ or 1 respectively, such that $Y_i = Y_i(W_i)$. The average treatment effect
is then defined as
\begin{equation*}
\tau \triangleq \mathbb{E}[Y_i(1) - Y_i(0)].
\end{equation*}
In order to identify $\tau$, we further assume unconfoundedness, i.e., that treatment assignment is as
good as random conditionally on the attributes $X_i$ \citep{rosenbaum_rubin_1983},
\begin{equation}
\label{eq:unconfoundedness}
\cb{Y_i(0), \, Y_i(1)} \indep W_i \cond X_i,
\end{equation}
and overlap, i.e., that the propensity score $e(\cdot)$ is bounded away from 0 and 1,
\begin{equation}
\label{eq:overlap}
e(x) \triangleq \PP{W_i = 1 \cond X_i = x}, \ \ \ \ \eta < e(x) < 1 - \eta,
\end{equation}
for all $x \in \RR^p$ and some $\eta > 0$.

In the case without any missingness in the attributes $X_i$, the problem of average treatment effect
estimation in the above setting is well understood. Several popular and consistent approaches to estimating
$\tau$ are built around the propensity score. The analyst first estimates the propensity score
$e(x)$ in  \eqref{eq:overlap}, and then estimates $\tau$ either via inverse-propensity weighting (IPW)
\begin{equation}
\label{eq:IPW}
\htau_{IPW} \triangleq \frac{1}{n} \sum_{i = 1}^n \p{\frac{W_i Y_i}{\he(X_i)} - \frac{(1 - W_i) Y_i}{1-\he(X_i)}},
\end{equation}
or by matching treated and control observations with similar values of the propensity score
\citep{abadie2016matching,rosenbaum_rubin_JASA1984,zubizarreta2012using}.

However, when the propensity score is somewhat difficult to estimate, methods that only
rely on the propensity score are in general dominated by bias due to estimation error in $e(\cdot)$,
and methods that also model the outcomes $Y_i$ can attain a better sample complexity;
see \citet{athey2018approximate}, \citet{chernozhukov_etal_2018} and \citet{van2011targeted}
for references and recent results. One particularly successful approach to combining these two
approaches to modeling is via augmented inverse-propensity weighting (AIPW) \citep{robins_etal_JASA1994},
\begin{equation}
\label{eq:AIPW}
\begin{split}
\htau_{AIPW} &\triangleq \frac{1}{n} \sum_{i = 1}^n \bigg(\hmu_{(1)}(X_i) - \hmu_{(0)}(X_i) \\
&\ \ \ \ \ \ \  + \frac{W_i }{\he(X_i)}\p{Y_i -\hmu_{(1)}(X_i)} - \frac{(1 - W_i)}{1-\he(X_i)}\p{Y_i -\hmu_{(0)}(X_i)}\bigg),
\end{split}
\end{equation}
where $\mu_{(w)}(x) \triangleq \EE{Y \cond X_i = x, \, W_i = w}$ and $\hmu_{(w)}(x)$ is an estimate thereof.
The AIPW estimator is often referred to as ``doubly robust'' because $\htau_{AIPW}$ is consistent for $\tau$ if either the
estimated outcome functions $\hmu_{(w)}(x)$ or the estimated propensity scores $\he(x)$ are consistent.

A key fact about doubly robust estimators as in \eqref{eq:AIPW} is that $\htau_{AIPW}$ can be $\sqrt{n}$-consistent
for $\tau$ and asymptotically Gaussian even in a non-parametric setting where \smash{$\hmu_{(w)}(\cdot)$}
and \smash{$\he(\cdot)$} are estimated, for instance using generic machine learning methods, at slower
non-parametric rates \citep{farrell2015robust}. In particular, provided use ``cross-fitting'', i.e.,
we do not use the $i$-th datapoint itself for making the predictions \smash{$\hmu_{(w)}(X_i)$} and \smash{$\he(X_i)$},
\smash{$\htau_{AIPW}$} using any choice of \smash{$\hmu_{(w)}(X_i)$} and \smash{$\he(X_i)$} attains $\sqrt{n}$ rates
of convergence whenever the product of the root-mean squared errors of \smash{$\hmu_{(w)}(X_i)$} and \smash{$\he(X_i)$}
decays faster than $1/\sqrt{n}$ \citep{chernozhukov_etal_2018,van2011targeted}.\footnote{Other methods, including those
based on inverse-weighting as in \eqref{eq:IPW}, can also sometimes achieve similarly good asymptotic performance,
but these results are generally more fragile and require considerably stronger regularity conditions than corresponding AIPW results
\citep{hirano2003efficient}.}

\section{Treatment Effect Estimation with Missing Attributes}

In this paper, we are interested in a more difficult variant of the above setting where the analyst cannot
always observe the full attribute vector. Rather, we assume that there is a ``mask''
$R_i \in \cb{1, \, \na}^p$ such that the analyst observes $X_i^* \triangleq R_i \odot X_i \in \cb{\RR \cup \na}^p$.
Here, $\odot$ denotes an element-wise product, such that $X_{ij}^* = X_{ij}$ if $R_{ij} = 1$ and
$X_{ij}^* = \na$ if $R_{ij} = \na$.\footnote{This representation of the incomplete data where the missing values are treated as a special category is chosen in view of the random forest approach handling this type of data.}

In current empirical practice, there are several approaches to treatment effect estimation with missing
attributes; but the literature studying this problem is rather scarce and most such approaches focus on IPW-form estimators as in \eqref{eq:IPW} \citep{rosenbaum_rubin_JASA1984, dagostino_rubin_JASA2000, seaman_white_2014, mattei_mealli_SMA2009, leyrat_etal_2019}.

The main contributions of this paper consist in (1) a dyadic classification of possible approaches to treatment effect estimation with missing attributes,
the first class relying on a variant of the unconfoundedness assumption while the second uses the classical missing values mechanism taxonomy;
(2) the proposal of two new estimators in the first class, a parametric and nonparametric estimator, both in an IPW and an AIPW form;
(3) the extension of previously introduced IPW estimators to the AIPW form in the second class; 
and (4) an extensive comparison of these estimators.
As preliminaries, below we review some paradigms for treatment effect estimation with missing
attributes.


\subsection{Unconfoundedness despite missingness} \label{sec:unconfounddespite}

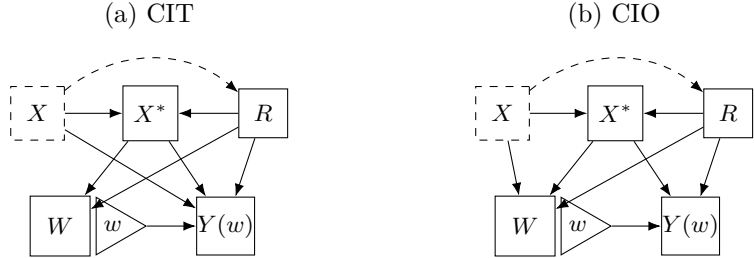
\begin{figure}
\caption{Causal graph depicting the assumptions \eqref{eq:citcio}.}
\label{fig:citcio_graph}
\begin{subfigure}[b]{0.48\textwidth}
\caption{CIT}
\centering
\begin{tikzpicture}

    \node[partial](X) at (0,2.5) {$X$};
    \node[regular polygon,regular polygon sides=4, draw, minimum width = 0.7 cm, inner sep=1pt](Xo) at (1.5,2.5) {$X^*$};
    \node[state](R) at (3,2.5) {$R$};
    
    \node[state] (W) at (0.3,1) {$W$};
    \node[intervene] (w) at (1,1) {\rotatebox{-30}{$w$}};
    \node [regular polygon,regular polygon sides=4, draw, minimum width = 0.7 cm, inner sep=-3pt] (Y) at (2.5,1) {$Y(w)$};    

    \path (X) edge (Y);
    \path (Xo) edge (Y);
    \path (Xo) edge (W);
    \path (w) edge (Y);
    
    \path (X) edge (Xo);
    \path (R) edge (Xo);
    \path (R) edge (W);
    \path (R) edge (Y);
    
    \path[dashed, bend  left=40] (X) edge (R); %
    
\end{tikzpicture}
\end{subfigure}
\begin{subfigure}[b]{0.48\textwidth}
\caption{CIO}
\centering
\begin{tikzpicture}
    
    \node[partial](X) at (0,2.5) {$X$};
    \node[regular polygon,regular polygon sides=4, draw, minimum width = 0.7 cm, inner sep=1pt](Xo) at (1.5,2.5) {$X^*$};
    \node[state](R) at (3,2.5) {$R$};
    
    \node[state] (W) at (0.3,1) {$W$};
    \node[intervene] (w) at (1,1) {\rotatebox{-30}{$w$}};
    \node [regular polygon,regular polygon sides=4, draw, minimum width = 0.7 cm, inner sep=-3pt] (Y) at (2.5,1) {$Y(w)$};    

    \path (X) edge (W);
    \path (Xo) edge (Y);
    \path (Xo) edge (W);
    \path (w) edge (Y);
    
    \path (X) edge (Xo);
    \path (R) edge (Xo);
    \path (R) edge (W);
    \path (R) edge (Y);
    
    \path[dashed, bend  left=40] (X) edge (R); %
\end{tikzpicture}
\end{subfigure}
\end{figure}

Perhaps the simplest way to work with missing attributes is to assume that the missingness
mechanism does not break unconfoundedness \eqref{eq:unconfoundedness}, i.e., that \citep{rosenbaum_rubin_JASA1984}
\begin{equation}
\label{eq:unconfoundedness-miss}
\cb{Y_i(0), Y_i(1)} \indep W_i \cond X_i^*.
\end{equation}
In this setting, \citet{dagostino_rubin_JASA2000} show that matching on the generalized propensity score
\begin{equation}
\label{eq:generalized-propensity}
e^*(x^*) \triangleq \PP{W_i = 1 \cond X_i^* = x^*}
\end{equation}
is consistent for $\tau$. In general, the simplest way to verify \eqref{eq:unconfoundedness-miss} is to
pair \eqref{eq:unconfoundedness} together with one of the two assumptions below
\citep{blake_etal_2019,mattei_mealli_SMA2009}
\begin{align}
\label{eq:citcio}
\begin{split}
& \left \{ 
\begin{array}{ll}
\text{CIT:}& \qquad W_i \independent X_i\,|\, X^*_i, R_i \\
\text{\textbf{or}} &\\
\text{CIO:}& \qquad Y_i(w) \independent X_i\,|\, X^*_i, R_i \quad \text{for }w\in\{0,1\},
\end{array}\right .
\end{split}
\end{align}
where CIT and CIO stand for \textit{conditional independence of treatment} and \textit{conditional independence of outcome} respectively.
Given these assumptions, \eqref{eq:unconfoundedness-miss} can be directly derived from the
causal graphs shown in Figure \ref{fig:citcio_graph} \citep{pearl1995causal,richardson_robins_2013}.

We note that fitting \eqref{eq:generalized-propensity} may appear difficult from the perspective of classical parametric
statistics; e.g., in order to run logistic regression, one needs to fit a separate parameter vector for each mask $r$.
However, many modern machine learning methods, including tree ensembles and neural networks, can readily handle
missing data and enable \eqref{eq:generalized-propensity} to be fit directly \citep{josse_etal_2019}.

\subsection{Missing values mechanisms}
\label{sec:mcar}

Another choice is to make assumptions about the missingness mechanism $R_i$. The most
popular approach is to take the missingness mechanism to be random (MAR) \citep{little_rubin_2002,rubin_1976},
i.e., for each possible mask $r \in \cb{1, \, \text{NA}}^p$,
\begin{equation}
\label{eq:MAR}
\mathbb{P}(R_i = r \cond X_i = x, \,W_i, \,Y_i) = \mathbb{P}(R_i = r \cond (X_i)_{r} = x_{r}, \,W_i,\, Y_i),
\end{equation}
where $X_{r}$ is the subset of entries of $X$ indexed by $\{j:\,r_j = 1\}$.
Under these assumptions, multiple imputation \citep{rubin_MINS1987, vanbuuren_2018} is a popular approach to treatment effect estimation
\citep{qu2009propensity,robins_wang_2000,rubin1978multiple,rubin2004multiple,seaman_white_2014}.
Under the condition that this imputation is ``proper'', i.e., that the missing attributes are simulated
from the correct conditional distribution, and correct model specification for the outcome and treatment, this method is consistent for IPW estimators \citep{seaman_white_2014}. 
Note that multiple imputation does not rely on the assumption \eqref{eq:unconfoundedness-miss} or the generalized propensity score, but it only requires the data to be MAR as in \eqref{eq:MAR}.

A stronger variant of the missing-at-random assumption \eqref{eq:MAR} is to assume
missingness to be completely at random (MCAR),
\begin{equation*}
\mathbb{P}(R_i = r \cond X_i, \,W_i,\, Y_i) = \mathbb{P}(R_i = r),
\end{equation*}
or equivalently
\begin{equation*}
R_i \indep \cb{X_i, \, Y_i, \, W_i}.
\end{equation*}
Under this assumption, further methods become available.
First, we can consistently estimate $\tau$ using only the subset of the
data with no missingness, i.e., $X_i = X_i^*$. Of course, using only a subset of the data results
in a loss of efficiency; however, this approach is simple and consistent. We emphasize that complete
case analysis is not valid under the weaker assumption \eqref{eq:MAR}; in that case, ignoring
observations with missingness will result in bias \citep{little_rubin_2002}.

Another algorithm that has been studied under the MCAR assumption is based on matrix completion
\citep{kallus_etal_2018}. Write $X$ and $X^*$ for the matrices with rows $X_i$ and $X_i^*$ respectively.
Then, assuming that $X$ is a potentially noisy realization of a low rank matrix $U$ and that unconfoundedness
\eqref{eq:unconfoundedness} holds with $X_i$ replaced by $U_i$, we can approximate $U$ from $X^*$ using methods for
low-rank matrix factorization \citep[e.g.,][]{candes2010matrix}, and then apply complete-data methods on the recovered $\hat{U}_i$. In cases where both MCAR and the low-rank
assumption hold, matrix factorization may be more efficient than complete case analysis and simpler than
multiple imputation.

\subsection{Discussion: The Traumabase study}
\label{sec:traumabase-assumptions}

In light of the previous discussion on the underlying (additional) assumptions required in the case of missing attributes, we argue that the Traumabase data is more likely to fall under the \textit{unconfoundedness despite missingness} assumption from Section \ref{sec:unconfounddespite} than the MAR assumption from Section \ref{sec:mcar}. Indeed, the administration of TA in the context of major trauma generally takes place under time pressure -- the more blood a patient looses, the more complications can occur -- and the medical staff cannot wait too long to collect a lot of information before deciding on the treatment. Therefore, if a value such as the evolution of the shock index level between arrival of the MICU\footnote{\textit{Mobile intensive care unit}, enhanced medical care team that takes care of the patient at the scene of the accident.} and arrival at the ICU, is not available because at least one measurement is missing -- for instance, due to transmission problems, the decision on the treatment will not depend on this feature. Another example could be information about the pre-hospital hemoglobin level: if the patient is in a severe state and immediate measures (such as resuscitation) are prioritized, then this measurement might not be made, however the consequently missing value is informative in the sense that it is due to the severe state of the patient, which might not necessarily be recorded explicitly in other observed features. These examples point in favor of the \textit{unconfoundedness despite missingness} assumption as they suggest that the missing values are not only missing for the analyst but have already been missing for the physician at the time of treatment administration. 

On the contrary, the MAR assumption seems plausible only for a subset of covariates. For instance, if the binary variable \textit{Cardiac.arrest.ph} indicates that the patient needed to be resuscitated, then this can explain the missing values for the blood pressure and heart rate during pre-hospital phase. And there are other incomplete variables such as the total quantity of volume expanders used in pre-hospital phase for which the missing values depend on several other recorded variables describing the need for volume expansion. But overall---due to the multitude of agents collecting the data in different circumstances and under important time constraints---such statements about the plausibility of MAR are difficult to assess on the whole of the registry.

\section{IPW and augmented IPW with Missing Attributes}
\label{sec:aipw-missing}

The previously discussed assumptions lead to two families of methods for treatment effect estimation with missing attributes. We now propose two IPW and AIPW estimators in the family derived from the \textit{unconfoundedness despite missingness} assumption (Section \ref{sec:unconfounddespite}). In the other family that relies on classical assumptions on the \textit{missingness mechanism} (Section \ref{sec:mcar}), we extend the existing multiple imputation IPW estimator to a doubly robust AIPW version. For the former family, we only present details for the AIPW estimators, their IPW counterparts can almost directly be read off the AIPW formulation below.

\subsection{Unconfoundedness despite missingness}
Under assumption \eqref{eq:unconfoundedness-miss}, the generalization to incomplete attributes is direct. First, estimate the
generalized propensity score $e^*(x^*)$ from \eqref{eq:generalized-propensity} and similarly
the generalized outcome model $\mu^*_{(w)}(x^*)$, and then form the AIPW estimator
\begin{equation}
\label{eq:AIPW-missing}
\begin{split}
&\htau_{AIPW^*} \triangleq \frac{1}{n} \sum_{i = 1}^n \bigg(\hmu^*_{(1)}(X_i^*) - \hmu^*_{(0)}(X_i^*) \\
&\ \ \ \  + \frac{W_i }{\he^*(X_i^*)}\p{Y_i -\hmu^*_{(1)}(X_i^*)} - \frac{(1 - W_i)}{1-\he^*(X_i^*)}\p{Y_i -\hmu^*_{(0)}(X_i^*)}\bigg).
\end{split}
\end{equation}
There are general results about AIPW that immediately guarantee that the above estimator $\htau_{AIPW^*}$ is
$\sqrt{n}$-consistent and asymptotically normal around $\tau$ given only weak regularity conditions
provided the product of the root-mean squared errors of the nuisance component estimates decay
as $o(n^{-1/2})$ \citep{chernozhukov_etal_2018}, and these results extend directly to the case where
the $X_i$ may contain missing values. Specifically, in order to get such results for $\htau_{AIPW^*}$,
it suffices to assume that
\begin{equation}
\label{eq:rate-prod}
\begin{split}
&\EE{\p{\frac{1}{\he^*(X_i^*)\p{1 - \he^*(X_i^*)}} -\frac{1}{e^*(X_i^*)\p{1 - e^*(X_i^*)}}}^2}^{\frac{1}{2}}\times \\
&\ \ \ \ \ \ \ \ \ \ \ \ \EE{\p{\hmu^*_{(W)}(X_i^*) - \mu^*_{(W)}(X_i^*)}^2}^{\frac{1}{2}}  = o\p{\frac{1}{\sqrt{n}}},
\end{split}
\end{equation}
i.e., that \smash{$\hmu^*_{w}(x^*)$} and \smash{$\he^*(x^*)$} are good approximations to the best
predictors we could have using on the partially observed predictors $x^*$.
Below, we instantiate the approach \eqref{eq:AIPW-missing} via both a parametric approach based on
logistic regression, and a non-parametric approach based on random forests.

\subsubsection{Parametric estimation of nuisance components}
\label{eq:param-EM}

For the parametric approach, we build on work by \citet{jiang_etal_2018} and \citet{schafer_1997} and logistic and linear forms
respectively for the generalized propensity score and outcome using the \emph{complete} covariates $x$.
The functions $\mu^*$ and $e^*$ that take in incomplete covariates $x^*$ are then estimated via EM \citep{dempster_etal_1977}.
The exact description of this parametric procedure for the AIPW estimator is outlined in Procedure \ref{box:proc-2a};
the resulting IPW and AIPW estimators will be denoted \smash{$\hat{\tau}_{EM}$}.

A major limitation of this approach is that, in order to justify use of the EM algorithm, one typically needs to make further assumptions
on the missing value mechanism; in particular, it is common to make the missing at random assumption \eqref{eq:MAR}.
In other words, although we did not require the missing at random assumption to identify $\tau$, this assumption is
used for consistent parametric estimation of $e^*(x^*)$ and $\mu^*_{(w)}(x^*)$. Below, we describe non-parametric alternative
that only needs the identifying assumption \eqref{eq:unconfoundedness-miss} to get consistency for $\tau$.

\begin{figure*}
\fbox{%
\parbox{\textwidth}{%
\textbf{Procedure 1:} \labeltext{1}{box:proc-2a} parametric AIPW with generalized propensity score and generalized response surfaces.
This algorithm provides an estimation for the average treatment effect $\tau$ via logistic and linear regressions, given incomplete covariates $X^{*}$, observed 
treatment assignment $W$ and outcome $Y$. We assume unconfoundedness despite missingness 
\eqref{eq:unconfoundedness-miss} and MAR \eqref{eq:MAR}. 

\begin{enumerate}
\item Fit a logistic model on $(W,X^*)$ using the stochastic approximation EM algorithm to obtain predictions for the generalized propensity score $e^*(X_i^*)$.
\item Fit two separate linear models on $(Y_{i:\,W_i=1},X^*_{i:\,W_i=1})$ and on $(Y_{i:\,W_i=0},X^*_{i:\,W_i=0})$ respectively via an EM algorithm to obtain predictions for $\mu^*_{(1)}(X_i^*)$ and $\mu^*_{(0)}(X_i^*)$ respectively.
\item Combine the predictions following \eqref{eq:AIPW-missing} to obtain a doubly robust estimation of $\tau$.
\end{enumerate}
}
}
\end{figure*}

\subsubsection{Non-parametric estimation of nuisance components}
\label{sec:nonparam}

As an alternative to fitting parametric models via EM as discussed above, one can also
directly estimate the functions \smash{$e^*(x^*)$} and \smash{$\mu^*_{(w)}(x^*)$} non-parametric. This
task may appear somewhat unusual, as the features $x^*$ take values in the augmented
space \smash{$\cb{\RR \cup \na}^p$}. However, many popular machine learning methods---including
decision trees, kernels and neural networks---can be adapted to this context, and standard arguments
still arguments for verifying consistency of these methods still apply \citep{josse_etal_2019}.
Then, once we have estimates of \smash{$e^*(x^*)$} and \smash{$\mu^*_{(w)}(x^*)$}, we can proceed 
to estimate the treatment effect using the AIPW estimator \eqref{eq:AIPW-missing} or
the analogous IPW estimator.

In this paper, we focus on non-parametric nuisance component estimation via (generalized)
random forests \citep{breiman2001random,athey_etal_AS2019}, with missing data handled
using the \textit{missing incorporated in attributes} (MIA) method of \citet{twala_etal_2008}.
The main idea of the MIA approach is give each split additional flexibility, such that missing
values may be sent on either side of the split independently of where the split occurred.
More specifically, as outlined by \citet{twala_etal_2008}, consider splitting on the $j$-th attribute
and assume that for some individuals, the value of $X_j$ is missing. MIA treats the missing values as
a separate category or code and the considers the following splits:
\begin{itemize}
\item $\{i: X_{ij} \leq t \text{ or } X_{ij} \text{ is missing}\}$ vs. $\{i:X_{ij} > t\}$
\item $\{i: X_{ij} \leq t\}$ vs. $\{i:X_{ij} > t \text{ or } X_{ij} \text{ is missing}\}$
\item $\{X_{ij}  \text{ is missing}\}$ vs. $\{X_{ij} \text{ is observed}\}$,
\end{itemize}
for some threshold $t$.
The MIA approach does not seek to model why some features are unobserved; instead, it
simply tries to use information about missingness to make the best possible splits for modeling
the desired outcome. Thus the MIA strategy work with arbitrary missingness mechanisms
and does not require the missing data to be MAR.\footnote{We conjecture that consistency proofs for random forests following, e.g., \citet{scornet2015consistency} or \citet{wager2015adaptive} extend to the case of MIA splitting and missing covariates. However, formal results of this type are not currently available.}

\sloppy{In order to estimate the average treatment effect, we use the estimator \eqref{eq:AIPW-missing}
with nuisance components extracted from a variant of the causal forests of \citet{athey_etal_AS2019}
that use MIA splitting to handle missing values.\footnote{We refer to Section 2.1 of \citet{athey2019estimating}
for a detailed discussion of how the doubly robust scores used in \eqref{eq:AIPW-missing} can be extracted
from a causal forest.} To do so, we have added the MIA splitting rule to the \texttt{causal\_forest} function in \texttt{grf}
\citep{grf}, and our proposed estimator can be computed by simply calling the function \texttt{average\_treatment\_effect}
on a trained causal forest.

\begin{figure*}
\fbox{%
\parbox{\textwidth}{%
\textbf{Procedure 2:} \labeltext{2}{box:proc-1a} nonparametric AIPW with generalized propensity score and generalized response surfaces.
This algorithm provides an estimation for the average treatment effect $\tau$ via random 
forests with MIA splitting rule, given incomplete covariates $X^{*}$, observed 
treatment assignment $W$ and outcome $Y$. We assume unconfoundedness despite missingness 
\eqref{eq:unconfoundedness-miss}. 

\begin{enumerate}
\item Train a causal forest on the potentially incomplete features $X^*$ using MIA splitting.
\item Extract out-of-bag estimates $\hmu^*_{(w)}(X_i^*)$ and $\he^*(X_i^*)$ from the causal forest.
\item Combine the predictions as in \eqref{eq:AIPW-missing} to obtain a doubly robust estimate $\hat{\tau}$ for $\tau$. 
\end{enumerate}
}
}
\end{figure*}

\subsection{Standard unconfoundedness and missingness mechanisms}
As discussed in  Section \ref{sec:mcar}, multiple imputation is a solution if the missingness mechanism is MAR as defined by \eqref{eq:MAR}. We propose to augment the multiple imputation approach to obtain an AIPW estimator: we proceed similarly to \citet{mattei_mealli_SMA2009}, i.e., we do multiple imputation using fully conditional equation (FCE) where we draw missing values from a joint distribution which is implicitly defined by the set of conditional distributions, proper imputation is ensured using a Bootstrap approach to reflect the sampling variability of the imputation  models parameters. Then, on each imputed data set $m\in\{1,\dots,\,M\}$, we compute an AIPW estimate $\hat{\tau}_{AIPW}^{(m)}$ given in \eqref{eq:AIPW} instead of the IPW estimate $\hat{\tau}_{IPW}^{(m)}$ given in \eqref{eq:IPW}. This approach is outlined in Procedure \ref{box:proc-3a}. We note that this method relies on the performance of the multiple imputation strategy; for instance in the case of FCE, the method requires correct specification of the conditional models which can be hard to assess in practice. We refer to \citet{carpenter_kenward_MIA2013} for a discussion on imputation strategies.



Another recent solution is based on matrix factorization \citep{kallus_etal_2018} as outlined in Procedure \ref{box:proc-4a} in the \ref{suppA}. Note that, unlike with multiple imputation, we only impute each datapoint once and consistency guarantees are only given under MCAR.

\begin{figure*}
\fbox{%
\parbox{\textwidth}{%
\textbf{Procedure 3:} \labeltext{3}{box:proc-3a} AIPW with multiple imputation.
This algorithm provides an estimation for the average treatment effect $\tau$ using multiple imputation, given incomplete covariates $X^{*}$, observed 
treatment assignment $W$ and outcome $Y$. We assume unconfoundedness 
\eqref{eq:unconfoundedness} and MAR \eqref{eq:MAR}. 

\begin{enumerate}
\item Choose number of imputations $M$, for instance $M=20$. Choose an imputation method. Impute the initial data $X^*$ using an $M$ times with the chosen imputation method to obtain $M$ complete data matrices $(X^{(1)}, \dots, X^{(M)})$.
\item  For every imputed data matrix $X^{(m)}$, $m\in\{1,\dots,M\}$: 
	 	\begin{description}
			\item [Option 1] Nonparametric regression.
			\begin{enumerate}
				\item Train a causal forest on the imputed features $X^{(m)}$.
				\item Extract out-of-bag estimates $\hmu_{(w)}(X_i^{(m)})$ and $\he(X_i^{(m)})$ from the causal forest.
				\item Combine the predictions following \eqref{eq:AIPW} to obtain a doubly robust estimation $\hat{\tau}$ for $\tau$. 
			\end{enumerate}
			\item [Option 2] Parametric regression (we additionally assume logistic-linear model specification for $(e,\mu_{(0)},\mu_{(1)})$).
			\begin{enumerate}
				\item Fit a logistic model to obtain predictions for the propensity score $e(X_i^{(m)})$ 
				\item Fit two separate linear models on $(Y_{i:\,W_i=1},X^{(m)}_{i:\,W_i=1})$ and on $(Y_{i:\,W_i=0},X^{(m)}_{i:\,W_i=0})$ respectively to obtain predictions for $\mu_{(1)}(X_i^{(m)})$ and $\mu_{(0)}(X_i^{(m)})$ respectively.
				\item Combine the predictions following \eqref{eq:AIPW} to obtain a doubly robust estimation $\hat{\tau}^{(m)}$ for $\tau$.
			\end{enumerate}
		\end{description}
\item Aggregate the $M$ estimations $(\hat{\tau}^{(1)},\dots,\hat{\tau}^{(M)})$: $\hat{\tau} = \frac{1}{M}\sum_{m=1}^M \hat{\tau}^{(m)}$.
\end{enumerate}
}
}
\end{figure*}

\section{Simulation study}
\label{sec:sim-study}
We assess the performance of the previously introduced treatment effect estimators in different scenarios, modifying the data generating process, the confounders' relationship structure, the unconfoundedness hypothesis, the missingness mechanism, the percentage of missing values, the sample size. The comparisons are twofold: (1) comparisons between IPW-baseline and AIPW-type estimators, (2) comparisons w.r.t. the assumptions on the underlying unconfoundedness and the missingness mechanism. 
Note that in all simulations, we only consider the well-specified case, i.e., we do not study the (parametric) estimators' performances in case of model mis-specification. More specifically, $e(x) = \sigma(\alpha_0 + \alpha^Tx + \epsilon_e)$ and $ \mu_{(w)}(x)= \beta_0 + \beta^Tx + w\tau + \epsilon_{\mu}$, where $\epsilon_e$ and $\epsilon_{\mu}$ are zero mean and independent noise terms. All simulations are implemented in R \citep{r-core}.\footnote{The code for reproducing the experiments presented in this work is available online at \url{https://github.com/imkemayer/causal-inference-missing}.}

\subsection{Methods overview}
We compare our approaches 
$\hat{\tau}_{EM}$ and  $\hat{\tau}_{MIA}$, denoted \textit{saem} and \textit{grf} in the experiments\footnote{These abbreviations refer to the algorithms used for the estimation of the nuisance parameters in the presence of missing values. For instance \textit{saem} stands for (stochastic approximation) EM algorithm.}, to the following methods, where we summarize their assumptions in Table \ref{tab:methods}:
\begin{itemize}
\item \textit{mice}: Procedure \ref{box:proc-3a} (and its IPW analogue detailed in the \ref{suppA}) with Option 2; we use the R package \texttt{mice} \citep{mice} and default options.
\item \textit{mf}: Procedure \ref{box:proc-4a} (and its IPW analogue detailed in the \ref{suppA}) with Option 2; we adapt the implementation\footnote{For details on the implementation of this last method, see \url{https://github.com/udellgroup/causal_mf_code}.} of \citet{kallus_etal_2018} based on the R package \texttt{softImpute} \citep{softImpute}. 
\item \textit{mean.loglin}: Imputation by the mean for the missing values and estimate $e$ with logistic regression on the mean imputed covariates and the two $\mu_{(w)}$ with two separate linear regressions.
\end{itemize}
For the parametric $\hat{\tau}_{EM}$ we use the R package \texttt{misaem} \citep{misaem}. We grow forests with missingness via the the MIA method; then, the estimator \eqref{eq:AIPW-missing} is implemented in the command \texttt{average\_treatment\_effect}.
Note that it is common to concatenate the initial or imputed data matrix $X$ and the binary mask $R$ for estimation or prediction and it is admitted that this addition can sometimes improve the analysis and generally does not deteriorate the result. Hence, in this work we only report results obtained by adding $R$.

In all cases, we consider inference using the bootstrap (i.e., we bootstrap the original data and repeat the whole process).


\begin{table}
\def\arraystretch{0.9}
\begin{tabularx}{\linewidth}{|@{ }p{1.3cm}|*{8}{>{\centering\arraybackslash}m{0.93cm}|}}
\cline{2-9}
 \multicolumn{1}{p{1.3cm}|}{} & \multicolumn{2}{>{\centering\arraybackslash}p{2.3cm}|}{Confounders \& Covariates} & \multicolumn{2}{>{\centering\arraybackslash}p{2.3cm}|}{Missingness} & \multicolumn{2}{@{\,}p{2.2cm}|}{Unconfoundedness} &  \multicolumn{2}{>{\centering\arraybackslash}p{2.2cm}|}{Models for $(W,Y)$} \\ 
\cline{2-9} 
\multicolumn{1}{p{1.3cm}|}{}  & \scriptsize \makecell{multiva- \\riate \\ normal} & \scriptsize \makecell{general} &  \multicolumn{1}{@{\,}p{0.93cm}|}{\scriptsize M(C)AR} & general & \scriptsize \eqref{eq:unconfoundedness} & \scriptsize \eqref{eq:unconfoundedness-miss} & \scriptsize logistic-linear & \scriptsize non-param. \\ 
\hline
 \scriptsize \textit{saem} & \cmark & \xmark & \cmark & \xmark & \xmark & \cmark & \cmark & \xmark \\ 
\hline 
 \scriptsize\textit{grf} & \cmark & \cmark & \cmark & \cmark & \xmark & \cmark & \cmark & \cmark \\ 
\hline
 \scriptsize\textit{mice} & \cmark & \cmark & \cmark & \xmark & \cmark & \cmark & \cmark & (\xmark) \\ 
\hline
 \scriptsize\textit{mf} & \cmark & \xmark & \cmark & \xmark & \cmark \space\space\space\scriptsize{(on $U$)} & \xmark & \cmark & (\xmark) \\ 
 \hline
 \scriptsize\textit{mean.loglin} & \xmark & \xmark & \xmark & \xmark & \xmark & \xmark & \xmark & \xmark \\ 
\hline
\end{tabularx}
\caption{Methods and their assumptions on the underlying data generating process. (\cmark indicates cases that can be handled by a method, whereas \xmark \,marks cases where a method is not applicable in theory; (\xmark) indicates cases without theoretical guarantees but with heuristic solutions.)}
\label{tab:methods}
\end{table}

\subsection{Data generation}
\label{sec:designs}

We  define different models for the generation of the confounders, covariates, missing values, treatment assignment and outcome.

\subsubsection{Confounders and covariates}

\paragraph{Model 1: Multivariate normally distributed confounders} We generate normally distributed confounders $X_{i\cdot} = [X_{i1}\;\; \dots \;\; X_{ip}]^T \sim \mathcal{N}(\mathbf{1},\Sigma)$, $i\in\{1,\dots, n\}$, for $p= 10$, where $\Sigma = I - 0.6\times(I-1)$, $\mathbf{X} = [X_{1\cdot} \,\dots\,X_{p\cdot}]^T \in\mathbb{R}^{n\times p}$.

\paragraph{Model 2: Latent classes model} We consider a Gaussian mixture model, i.e., we first generate class labels $C$ from a multinomial distribution with three categories. Then the confounders of observation $i$, $X_{i\cdot}$, are sampled from the corresponding class distribution, i.e., $X_{i\cdot}\sim\mathcal{N}(\mu(c_i), \Sigma(c_i)) \,|\, C_i = c_i$. 

Treatment and outcome are defined using the logistic-linear model in the following way: we define $\logit(e^*(X_{i\cdot}^*)) = (\alpha(C_i))^TX_{i\cdot}^*$. This allows us to add an additional interaction between treatment and the latent class. Analogously, the outcome is defined as $Y_i \sim\mathcal{N}((\beta(C_i))^TX_{i\cdot}^* + \tau W_i, \sigma^2)$.

\paragraph{Model 3: Low rank matrix factorization} We adapt the simulation framework from \citet{kallus_etal_2018} by generating $U_{i\cdot} = [U_{i1}\;\; \dots \;\; U_{id}]^T \sim \mathcal{N}(0,I_d)$ and defining $X = UV^T$ for some fixed matrix $V\in\mathbb{R}^{p\times d}$, with $d=3$. 

\paragraph{Model 4: Hierarchical data-generating model} An alternative to defining a Gaussian mixture model, is to use a simplified shallow version of a \textit{deep latent variable model} (DLVM, \citet{kingma_welling_2014}): the codes $C$ are sampled from a normal distribution $\mathcal{N}_d(0,1)$. Covariates $X_i$ are then sampled from $\mathcal{N}_p(\mu(c),\Sigma(c))\,|\, C_i=c$, where $$(\mu(c),\Sigma(c)) = (V\tanh(Wc+a)+b,
\exp(\gamma^T(Wc+a) + \delta)I_p),$$ 
and the weights in $V\in\mathbb{R}^{p\times 5}$ and $W\in\mathbb{R}^{5\times d}$ are respectively sampled from a standard normal and a uniform distribution (and similarly for the offsets $a$ and $b$). We fix $d=3$. Results for this model are reported in the \ref{suppA}.

\subsubsection{Missing values}

We generate missing values either under MCAR (i.e., $\mathbb{P}(R_{ij}=1) = 1-\mathcal{B}(\eta)$ such that on average we have $\eta n p$ missing values) or as informative\footnote{By informative we designate all non-ignorable missingness mechanisms, where the probability of observing missing values depends on the missing values.} missing values (missing values in $X_{\cdot, 1:5}$ are generated depending on the quantiles of $X_{\cdot, 1:5}$ such that there are about $\eta n p/2$ missing values). In the results presented here we fix $\eta = 0.3$. 

\subsubsection{Treatment assignment and outcome}

For models 1, 3 and 4, treatment assignment and outcome are defined under either of the unconfoundedness assumptions.
\paragraph{Unconfoundedness despite missingness} We define $\logit(e^*(X_{i\cdot}^*)) = \alpha_0 + \alpha^TX_{i\cdot}^*$. Analogously, the outcome is defined as $Y_i \sim\mathcal{N}(\beta_0 + \beta^TX_{i\cdot}^* + \tau W_i, \sigma^2)$.
\paragraph{Complete data unconfoundedness} We define $\logit(e(X_{i\cdot})) = \alpha_0 + \alpha^TX_{i\cdot}$. Analogously, the outcome is defined as $Y_i \sim\mathcal{N}(\beta_0 + \beta^TX_{i\cdot} + \tau W_i, \sigma^2)$.

For model 2, treatment assignment and outcome are defined under unconfoundedness on the latent factors $U$ as follows: $\logit(e(U_{i\cdot})) = \alpha_0 + \alpha^TU_{i\cdot}$. Analogously, the outcome is defined as $Y_i \sim\mathcal{N}(\beta_0 + \beta^TU_{i\cdot} + \tau W_i, \sigma^2)$

We refer to the \ref{suppA} for details on how to simulate treatment and outcome under assumption \eqref{eq:unconfoundedness-miss} (or rather \eqref{eq:unconfoundedness} and \eqref{eq:citcio}).

\subsection{Results}
\label{sec:results}
We report the estimations for a fixed average treatment effect using the previously described estimation methods. All figures in this study are generated from 100 simulations for sample sizes $n\in\{100, 500,1000, 5000\}$, we fix the proportion of missing values at 30\% throughout all experiments; and the true treatment effect $\tau$ is reported as black solid line. The \textit{standard unconfoundedness} setting corresponds to assumption \eqref{eq:unconfoundedness}, while \textit{unconfoundedness despite missingness} corresponds to \eqref{eq:unconfoundedness-miss}.

\begin{figure}[h!]
     \centering
     \begin{subfigure}[b]{0.83\textwidth}
         \centering
         \includegraphics[width=0.9\textwidth]{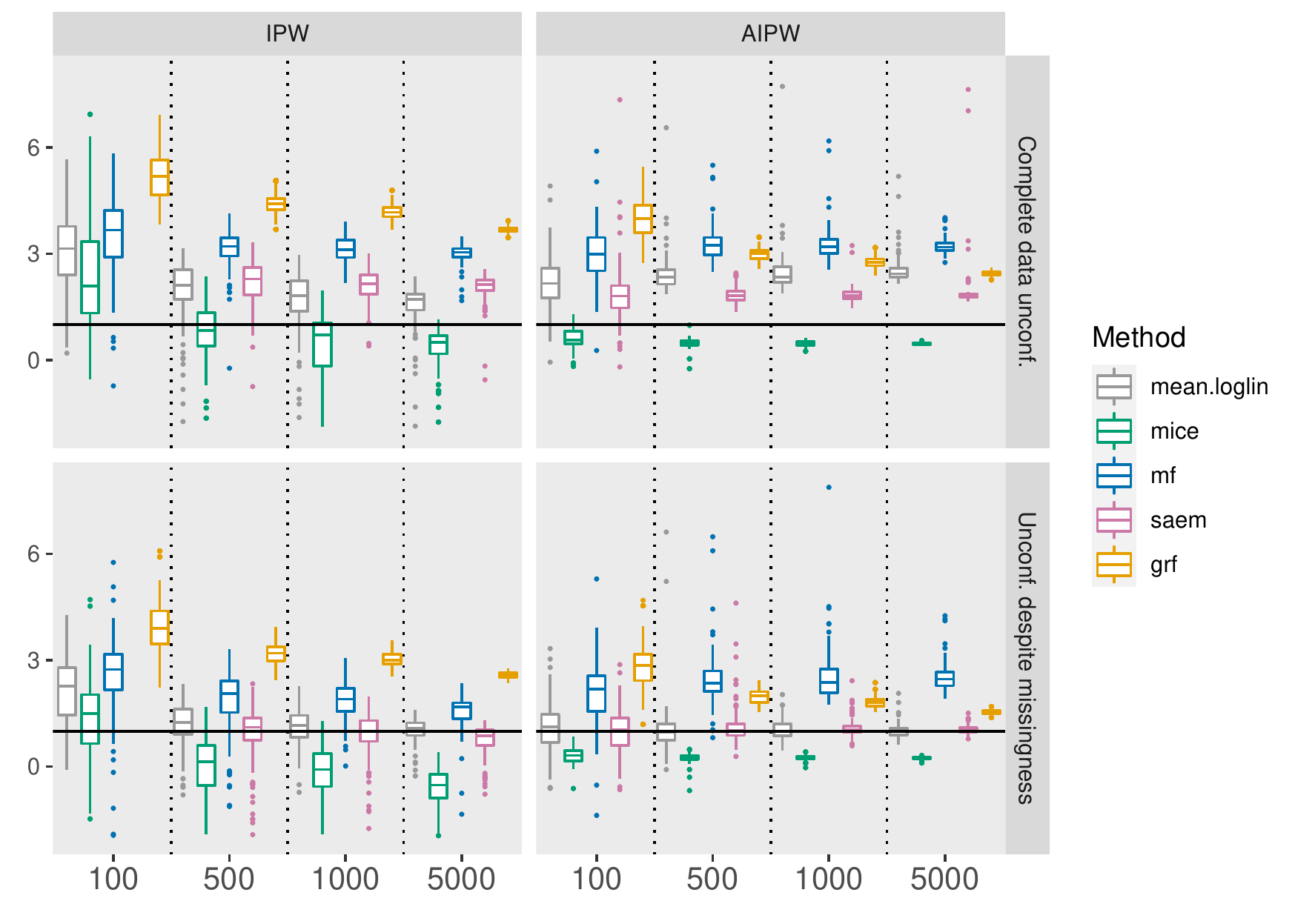}
         \caption{MCAR (with 30\% missing values in $X_{\cdot,1:10}$)}
         \label{fig:simulations_set2_mcar}  
     \end{subfigure}
     \par\medskip
     \begin{subfigure}[b]{0.83\textwidth}
         \centering
         \includegraphics[width=0.9\textwidth]{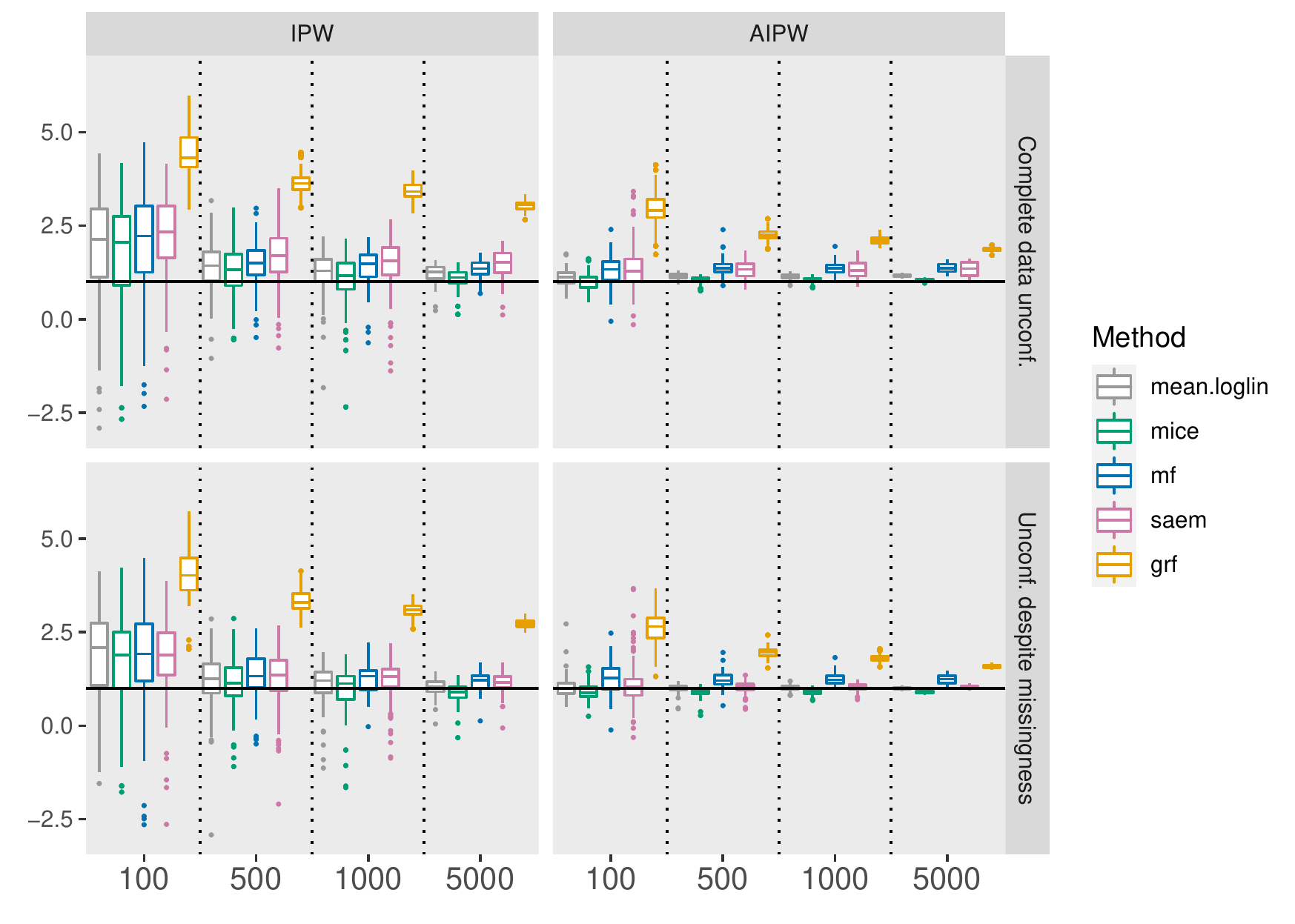}
         \caption{Informative missing values (with 30\% missing values in $X_{\cdot,1:5}$)}
         \label{fig:simulations_set2_mnar}
     \end{subfigure}
        \caption{Model 1. IPW and AIPW estimations across simulation designs described in Section \ref{sec:designs}. We report results for all combinations of $n\in \{100, 500, 1000, 5000\}$, missing values mechanism $\in \{\text{MCAR, general}\}$ and unconfoundedness $\in\{\cdot\text{ despite missingness, complete data }\cdot\}$. Results are displayed for 100 runs of every setting.}
\label{fig:simulations_param}
\end{figure}

\begin{figure}[h!]
     \centering
     \begin{subfigure}[b]{0.83\textwidth}
         \centering
         \includegraphics[width=0.9\textwidth]{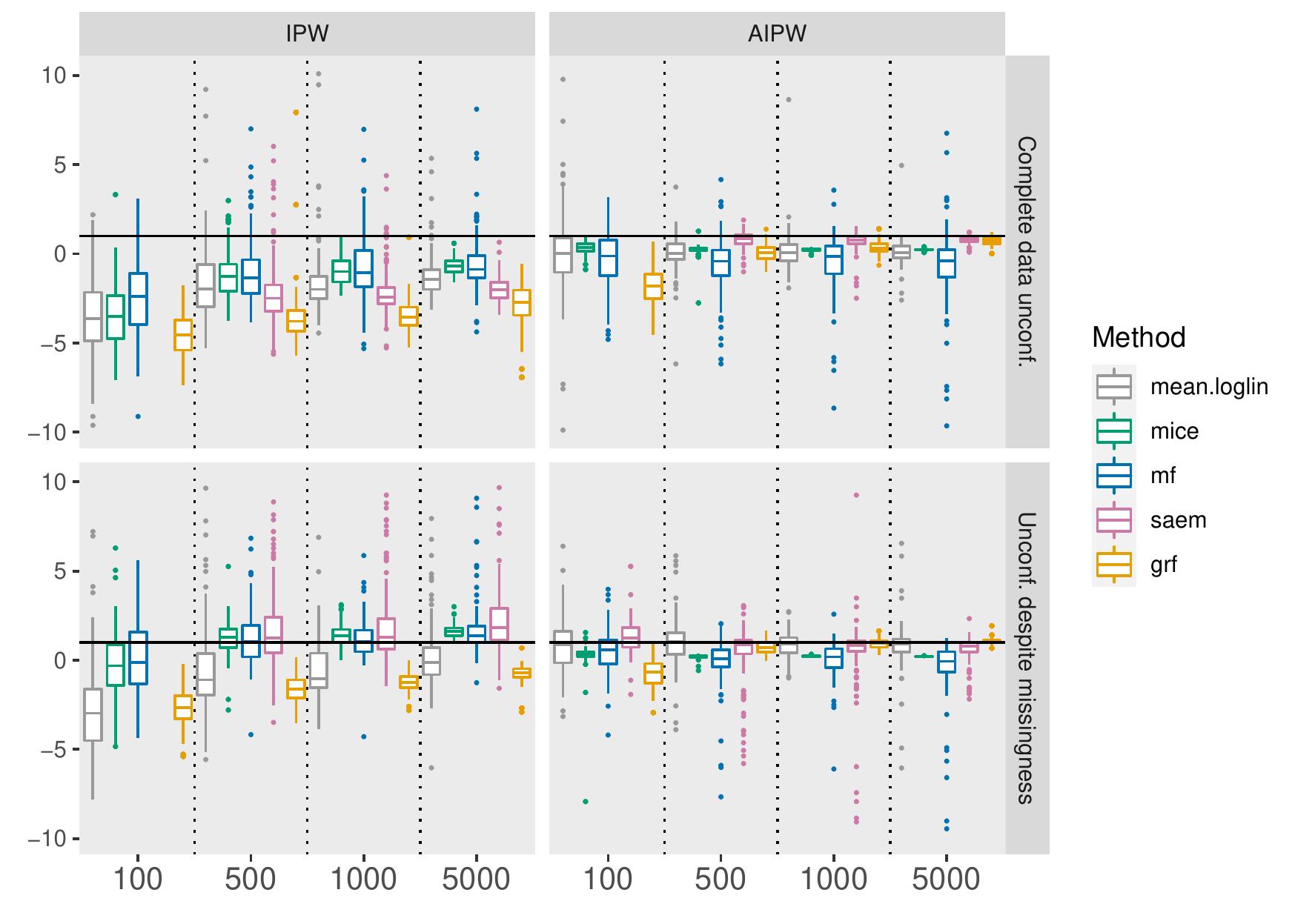}
         \caption{MCAR (with 30\% missing values in $X_{\cdot,1:10}$)}
         \label{fig:simulations_latent_mcar}
     \end{subfigure}
     \par\bigskip
     \begin{subfigure}[b]{0.83\textwidth}
         \centering
         \includegraphics[width=0.9\textwidth]{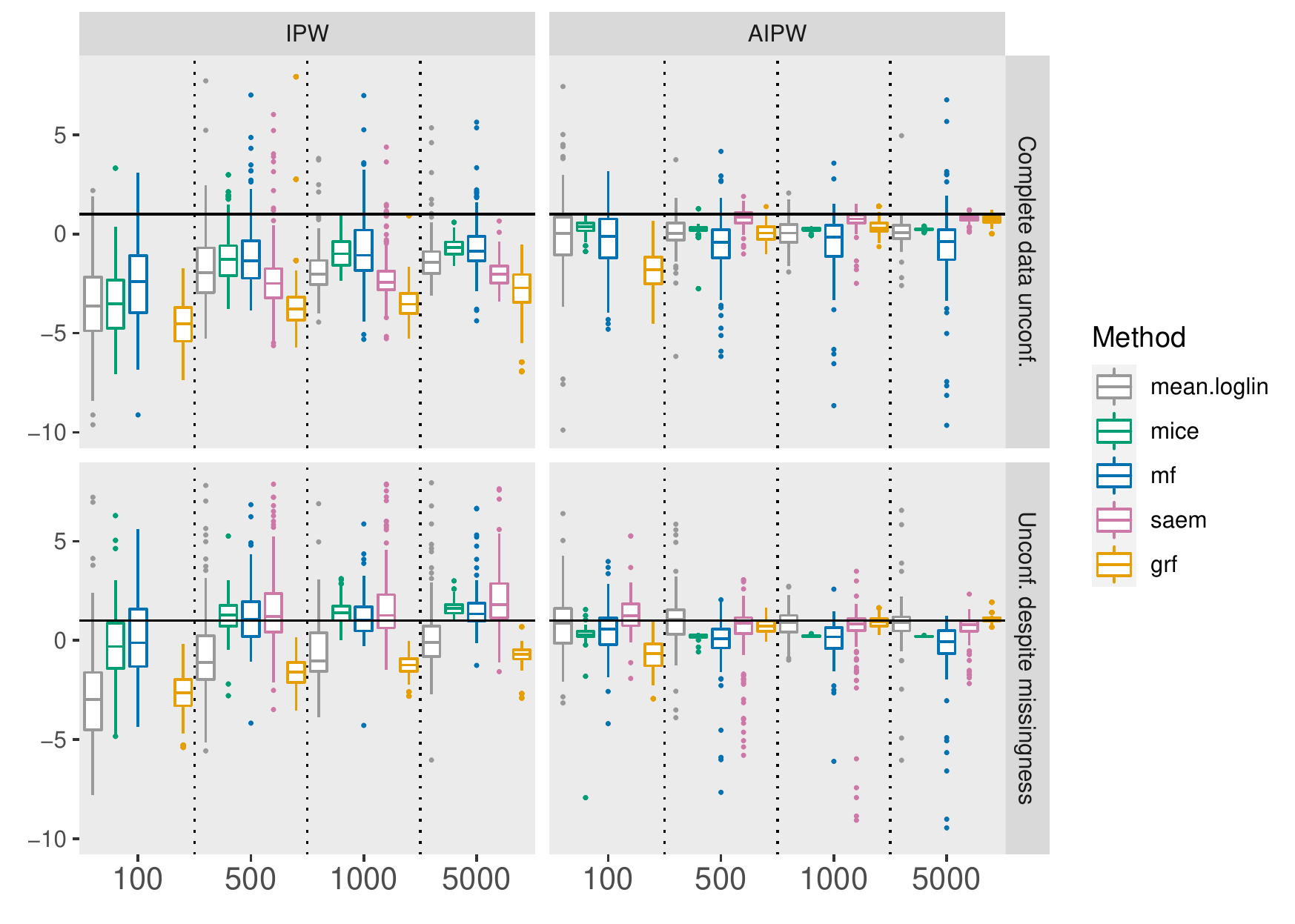}
         \caption{Informative missing values (with 30\% missing values in $X_{\cdot,1:5}$)}
         \label{fig:simulations_latent_mnar}
     \end{subfigure}
        \caption{Model 2. IPW and AIPW estimations across simulation designs described in Section \ref{sec:designs}. We report results for all combinations of $n\in \{100, 500, 1000, 5000\}$, missing values mechanism $\in \{\text{MCAR, general}\}$ and unconfoundedness $\in\{\cdot\text{ despite missingness, complete data }\cdot\}$. Results are displayed for 100 runs of every setting.}
        \label{fig:simulations-nonlinear}
\end{figure}

\begin{figure}[h!]
     \centering
     \begin{subfigure}[b]{0.8\textwidth}
         \centering
         \includegraphics[width=0.9\textwidth]{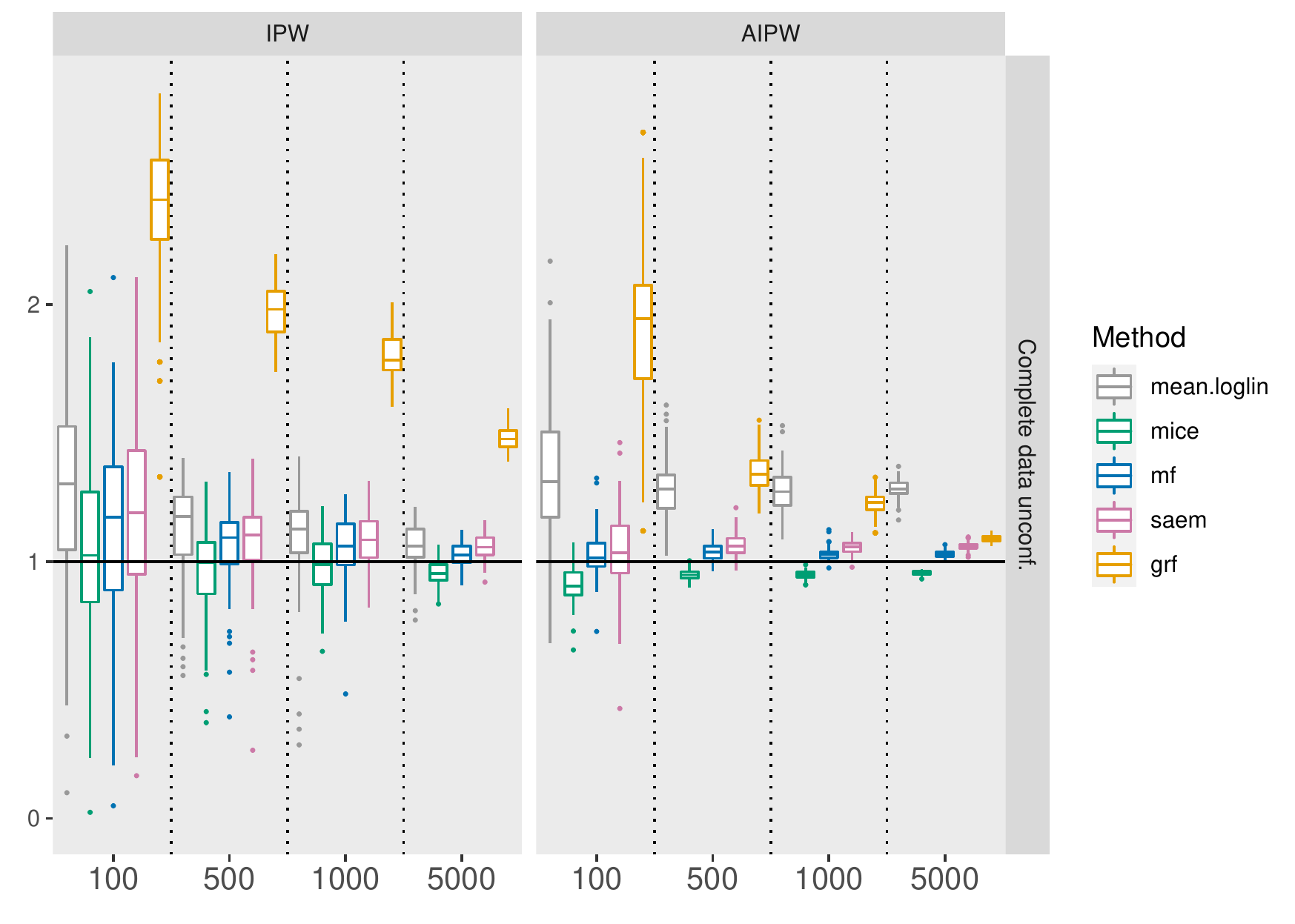}
         \caption{MCAR (with 30\% missing values in $X_{\cdot,1:10}$)}
         \label{fig:simulations_set3_mcar}
     \end{subfigure}
     \begin{subfigure}[b]{0.8\textwidth}
         \centering
         \includegraphics[width=0.9\textwidth]{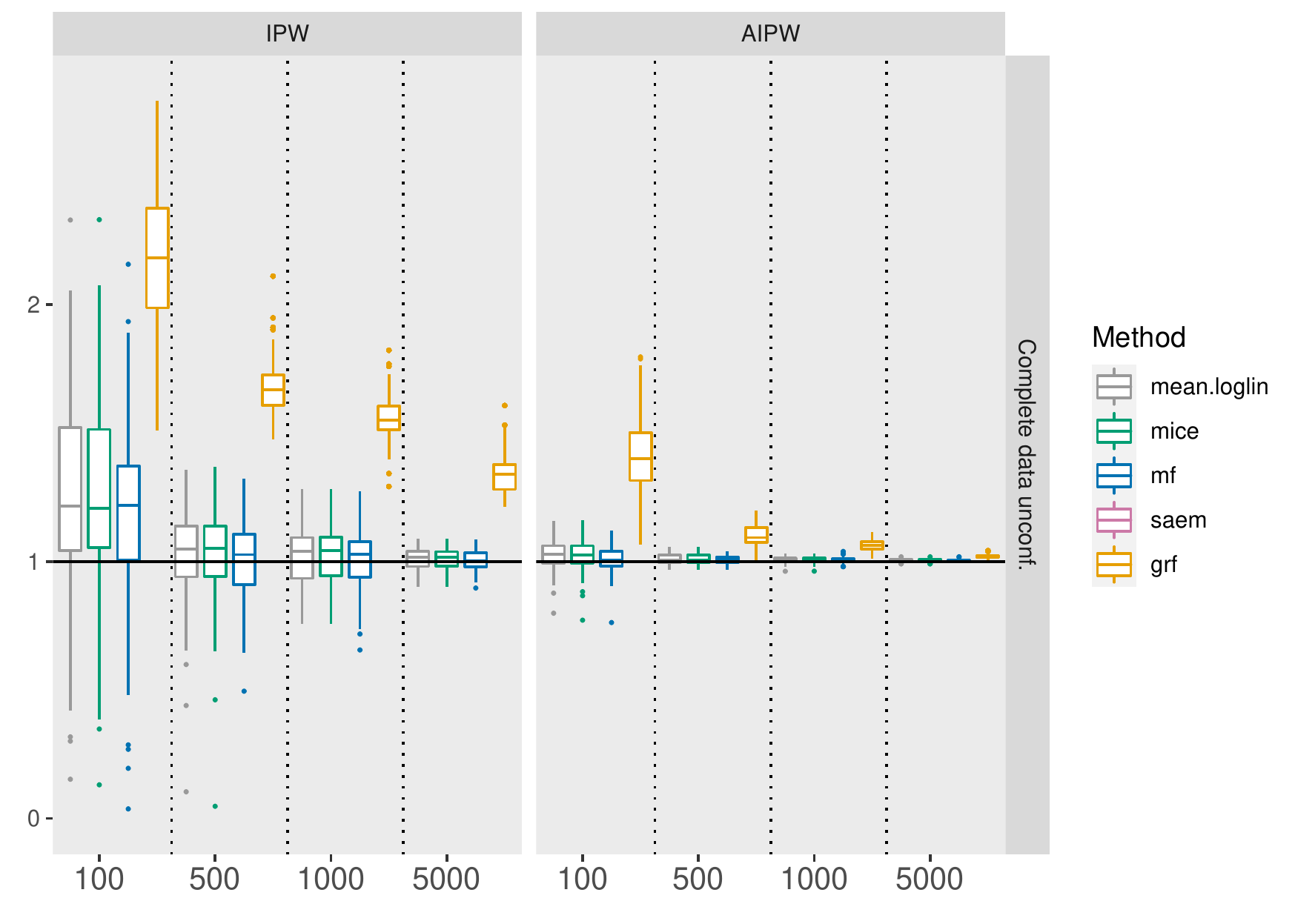}
         \caption{Informative missing values (with 30\% missing values in $X_{\cdot,1:5}$)}
         \label{fig:simulations_set3_mnar}
     \end{subfigure}
        \caption{Model 3. IPW and AIPW estimations across simulation designs described in Section \ref{sec:designs}. We report results for all combinations of $n\in \{100, 500, 1000, 5000\}$ and missing values mechanism $\in \{\text{MCAR,\, general}\}$. Results are displayed for 100 runs of every setting.}
        \label{fig:simulations-nonlinear-set3}
\end{figure}


\subsection{Take-home message from the simulation study}
The results from this first simulation study can be summarized in several general observations:
\begin{itemize}
\item Augmented IPW outperform their IPW equivalents throughout all scenarios (both in terms of variability and of bias), this behavior is analogous to the behavior in the well understood complete data setting.
\item All methods perform well if their assumptions on the underlying data generating process are met (see Table \ref{tab:methods}).
\item For multiple imputation (\textit{mice}) there is a small remaining bias, even for large sample sizes. In some cases, when the assumptions for this method are met, based on the theorem from \citet{seaman_white_2014} on multiple imputation with $M=\infty$ imputations, it is expected that an increase of the number of imputations should decrease this remaining bias in these cases.
\item The tree-based estimation using the MIA splitting rule (\textit{grf}) generally performs at least as well as multiple imputation but yields unbiased results if ``unconfoundedness despite missingness" \eqref{eq:unconfoundedness-miss} holds.
\item Mean imputation coupled with concatenation of the imputed data with the mask and parametric estimation empirically performs well, provided that \eqref{eq:unconfoundedness-miss} holds. However, the concatenation of the mask $R$ appears necessary, since otherwise this approach is biased as soon as \eqref{eq:unconfoundedness-miss} is violated, and in this case it is outperformed by competing methods.
\item The EM-based estimator (\textit{saem}) performs well under correct specification (multivariate Gaussian confounders, logistic treatment assignment, linear outcome, M(C)AR missing data mechanism, \eqref{eq:unconfoundedness-miss} satisfied) and adding the mask to the initial data matrix yields unbiased estimates even if the missing data mechanism is not ignorable. It fails however in the cases where the data is not i.i.d. Gaussian.
\end{itemize}

In conclusion, the type of unconfoundedness assumption is important for the choice of the estimation strategy. Once the type is fixed, the choices between simple and doubly robust and between parametric and non-parametric estimation depend on the \textit{a priori} on the data generating processes. However, in general, we recommend privileging the doubly robust strategy.

For a more detailed discussion of the simulation results, we refer to the \ref{suppA}.

\section{Application on observational critical care management data}
\label{sec:application}
As announced in the introduction we apply our methods to clinical data from a French observational database on major trauma patients. The medical question we aim to answer is whether administrating the drug TA has an effect on in-ICU mortality for patients with traumatic brain injury. 

\subsection{Data and causal DAG}
For our analysis we used 20,037 currently available validated patient records, validated by the medical expert team after a first pre-treatment. The pre-treatment consisted in identifying outliers clearly due to erroneous inputs and recoding missing values that are not really missing (for instance the variable informing previous pregnancies is evidently consistently missing, or ideally set to false, for male patients, etc.).\footnote{The code for pre-treatment and for estimating the treatment effect on this data are available at \url{https://github.com/imkemayer/causal-inference-missing}.} Out of these 20,047 patients, 8,269 are identified as having a traumatic brain injury (defined by the medical expert team as either the presence of a brain lesion visible on the first computed tomography (CT) scan---which is generally taken within the first three hours after the accident---or as a head AIS score\footnote{The head Abbreviated Injury Score indicates, on a scale from one to six, the severity of the most severe observed brain lesion. This score is defined in the context of the Abbreviated Injury Scale proposed by the American Association for Automotive Medicine. See the \ref{suppA} or \url{https://www.aaam.org/abbreviated-injury-scale-ais/} for more information.} greater or equal 2). Additionally, we excluded a total of 21 patients among this group coming from Trauma centers with too few observations, having joined the registry group several years after the majority of all other Trauma centers.

The treatment of interest, TA, is an antifibrinolytic agent limiting excessive bleeding and it is currently used in patients suspected of developing an hemorrhagic shock, a state in which the body is no longer able to provide vital organs with sufficient quantities of dioxygen to sustain them. The average cost of a dose of TA lies below 10\texteuro $\,$ and the drug is generally available immediately after the arrival of the medical first responders team at the place of the accident. 
It is now recommended to administer this drug to patients at risk of developing an hemorrhagic shock. 

In order to clarify the previously raised causal question given the data, we first establish a causal graph in order to summarize the a priori on existing confounding and to highlight the causal question, as suggested, for instance, by \citet{lederer_etal_2019, blake_etal_2019}. The causal graph in Figure \ref{fig:dagitty} is the result of a two-step Delphi procedure in which  six anesthetists and resuscitators specialized in critical care first selected covariates related to either treatment or outcome or both and second classified these covariates into confounders and predictors of only treatment or outcome. 
The absence of an exact timestamp for the drug administration is compensated by the fact that it is always given within the first three hours from the accident and that the treatment does not have an immediate effect on variables such as blood pressure, hemoglobin level or the Glasgow Coma Scale (GCS) which are measured at various moments within the first three hours.

\begin{figure}[h!]
\begin{center}
\includegraphics[width=0.75\textwidth]{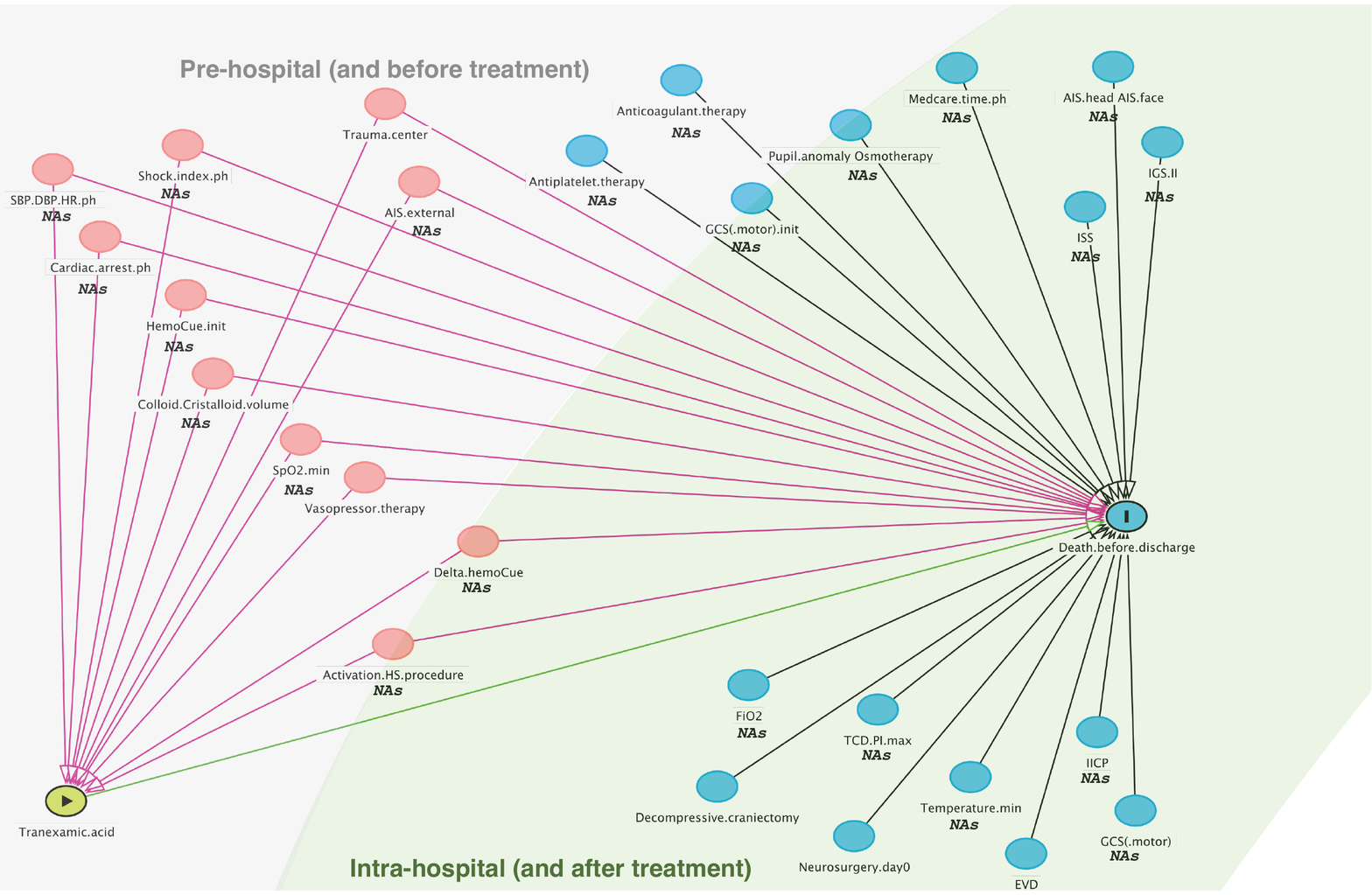}
\caption{Causal graph representing treatment, outcome, confounders and other predictors of outcome (Figure generated using DAGitty \citep{textor_etal_2011}; \texttt{NAs} indicates variables that still have missing values after pre-treatment).}
\label{fig:dagitty}
\end{center}
\end{figure}

From this graph it becomes clear as well that a method that incorporates a model of the outcome as a function of the identified potential predictors (red and blue vertices in the graph) might achieve more precise results than a method that uses the observed outcome directly. The large number of predictors of the outcome is due both to the medical complexity of traumatic brain injury and to the ambiguous treatment target: the assignment is made in the context of hemorrhagic shock but recently there is some evidence that there might also be a beneficial effect in the context of traumatic brain injury \citep{hijazi_etal_2015}.

\subsection{Results}

First, we recall the estimand we aim at estimating in this context: we are interested in the average treatment effect of the treatment on mortality among traumatic brain injury patients.
When adjusting for confounding using the identified confounders (red nodes on the graph in Figure \ref{fig:dagitty}), using additional predictors for the outcome model (blue nodes on the graph in Figure \ref{fig:dagitty}), we obtain the following estimations in Figure \ref{fig:dr-traumabase} of the direct causal effect of TA on in-ICU mortality among traumatic brain injury patients. 

\begin{figure}[h!]
\begin{center}
\includegraphics[width=0.6\textwidth]{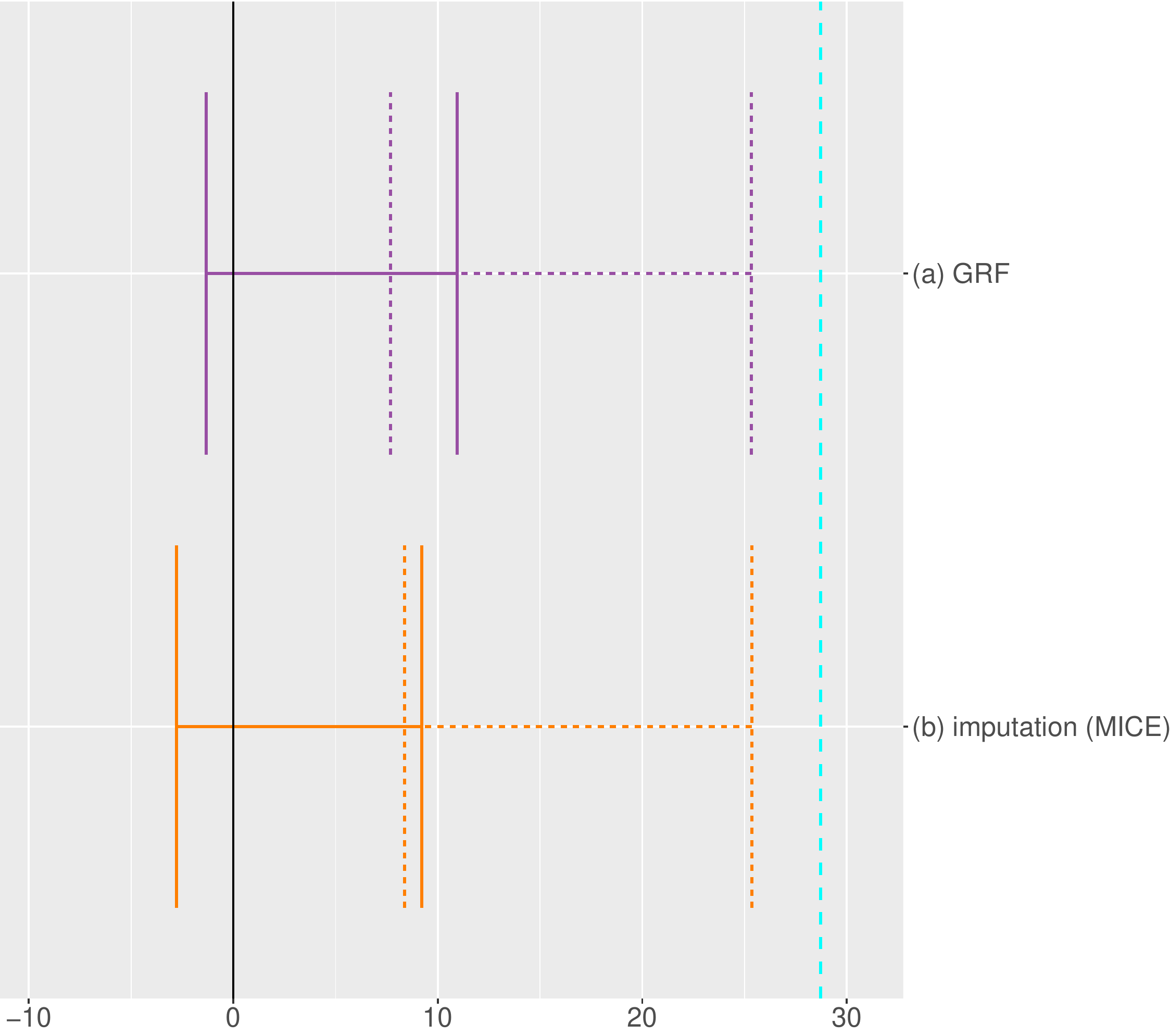}
\caption[ATE estimations on Traumabase data (solid: doubly robust estimates; dotted: IPW estimates; dashed vertical line: without adjustment; $x$-axis: $\hat{\tau}$ and asymptotically valid confidence intervals]{ATE estimations on Traumabase data (solid: doubly robust estimates; dotted: IPW estimates; dashed vertical line: without adjustment; $x$-axis: $\hat{\tau}$ and bootstrap confidence intervals\footnotemark). \textit{Note:} Positive ATE $\equiv$ increase of mortality.}
\label{fig:dr-traumabase}
\end{center}
\end{figure}

\footnotetext{Values on the $x$-axis are multiplied by 100 for better readability. The results can be read as difference in percentage points between mortality rate in the treatment groups.}




Unlike the simulations of the previous paragraph, the real-world medical data is more complicated and some concessions have to be made to apply the previously discussed method. For instance, due to an important number of outliers in the variable \textit{Medcare.time.ph} that are related with inconsistent units of the recorded values and with patient transfers from one hospital to another, we chose to drop this variable in our analyses since, according to the practitioners, its predictive power does not outweigh the potential issues related to inconsistent recording of this variable.

Note that apart from the issue with the variable \textit{Medcare.time.ph}, the estimation via random forest with MIA splitting rule does not require any pre-processing of the data and is therefore straightforward when using the \textit{grf} package. 

Here, we only consider three pairs of methods: \textit{grf} and \textit{mice}. We do not test \textit{saem} and \textit{mf} since currently both these methods have not been derived for heterogeneous data.\footnote{Concatening the mask with the data matrix does not lead to major changes in the estimations, therefore we only report results obtained when including the mask.} 
A first observation on the results reported in Figure \ref{fig:dr-traumabase} is the concordance of the two estimators: none of the AIPW-type estimation strategies allows to reject the null hypothesis of no treatment effect. As discussed in Section \ref{sec:traumabase-assumptions}, it can be argued which family of methods has more plausible underlying assumptions on the Traumabase data, but in our opinion the \textit{unconfoundedness despite missingness}---and therefore the \textit{grf} estimations---are most suited for our specific application. When comparing covariate balance for both methods in terms of standardized mean differences, we note that both methods achieve similar balance on the observed values (see results reported in the \ref{suppA}) but, as expected, only GRF additionally achieves balance on the response pattern (Figure \ref{fig:balanceNA}). Since there is consensus by the medical experts that certain missing values are not missing at random, achieving balance on the response pattern is a relevant feature for interpreting the estimation results. A remaining issue might consist in the overlap assumption which is generally difficult to assess in most medical applications and which might be slightly violated due in part to the heterogeneity of patient profiles and it could be argued that for certain patients the probability of receiving the treatment is zero. However, the lack of a standardized protocol for tranexamic acid administration favors the overlap assumption even for this group of patients. A solution to handle weak overlap is the use of overlap weights \citep{li_etal_JASA2018} and we give the results using this alternative to inverse propensity weights in the \ref{suppA}.

\begin{figure}[h!]
     \centering
     \begin{subfigure}[b]{0.44\textwidth}
         \centering
         \includegraphics[width=\textwidth]{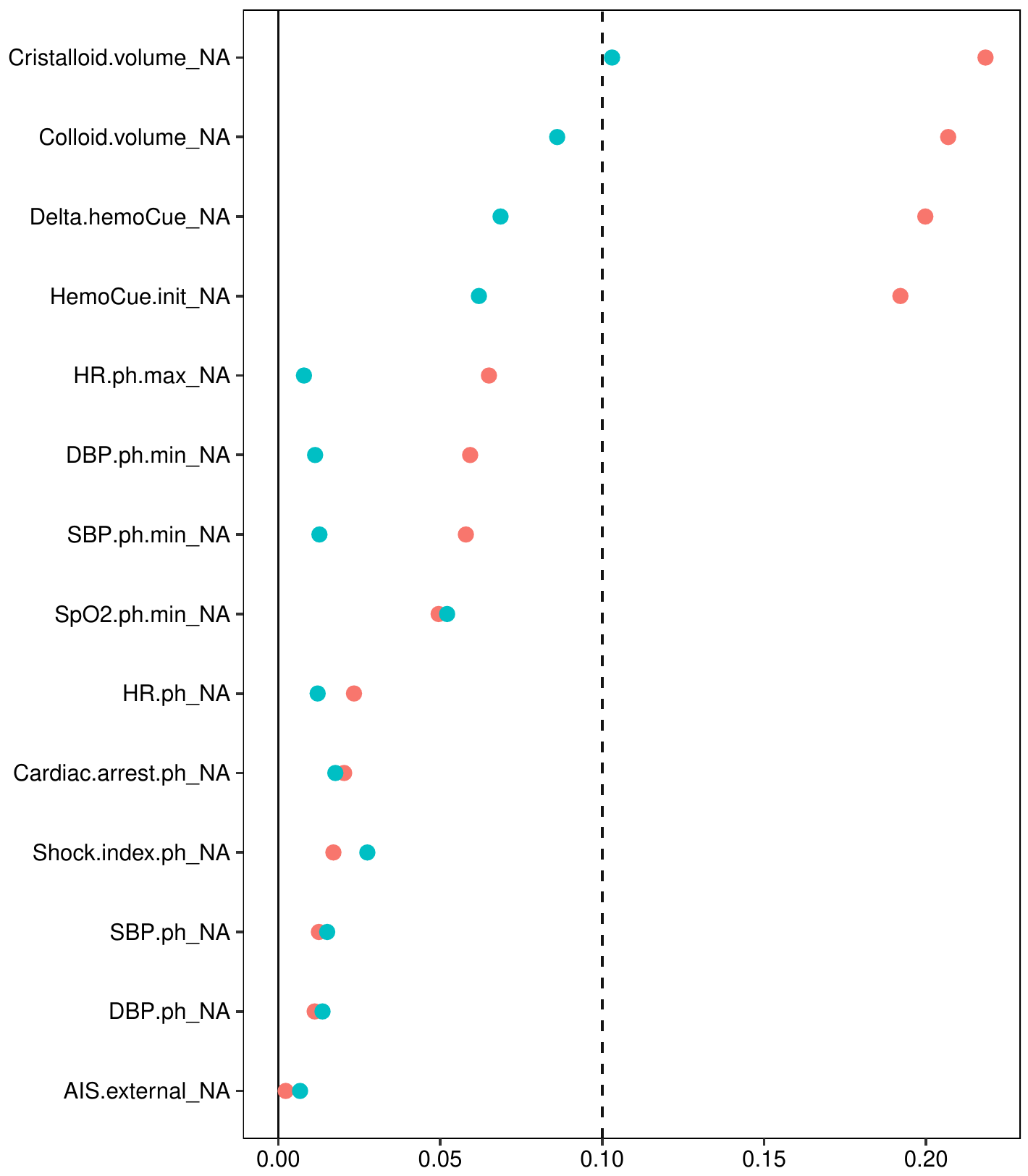}
         \caption{GRF}
         \label{fig:grf-balanceNA}
     \end{subfigure}
     \begin{subfigure}[b]{0.44\textwidth}
         \centering
         \includegraphics[width=\textwidth]{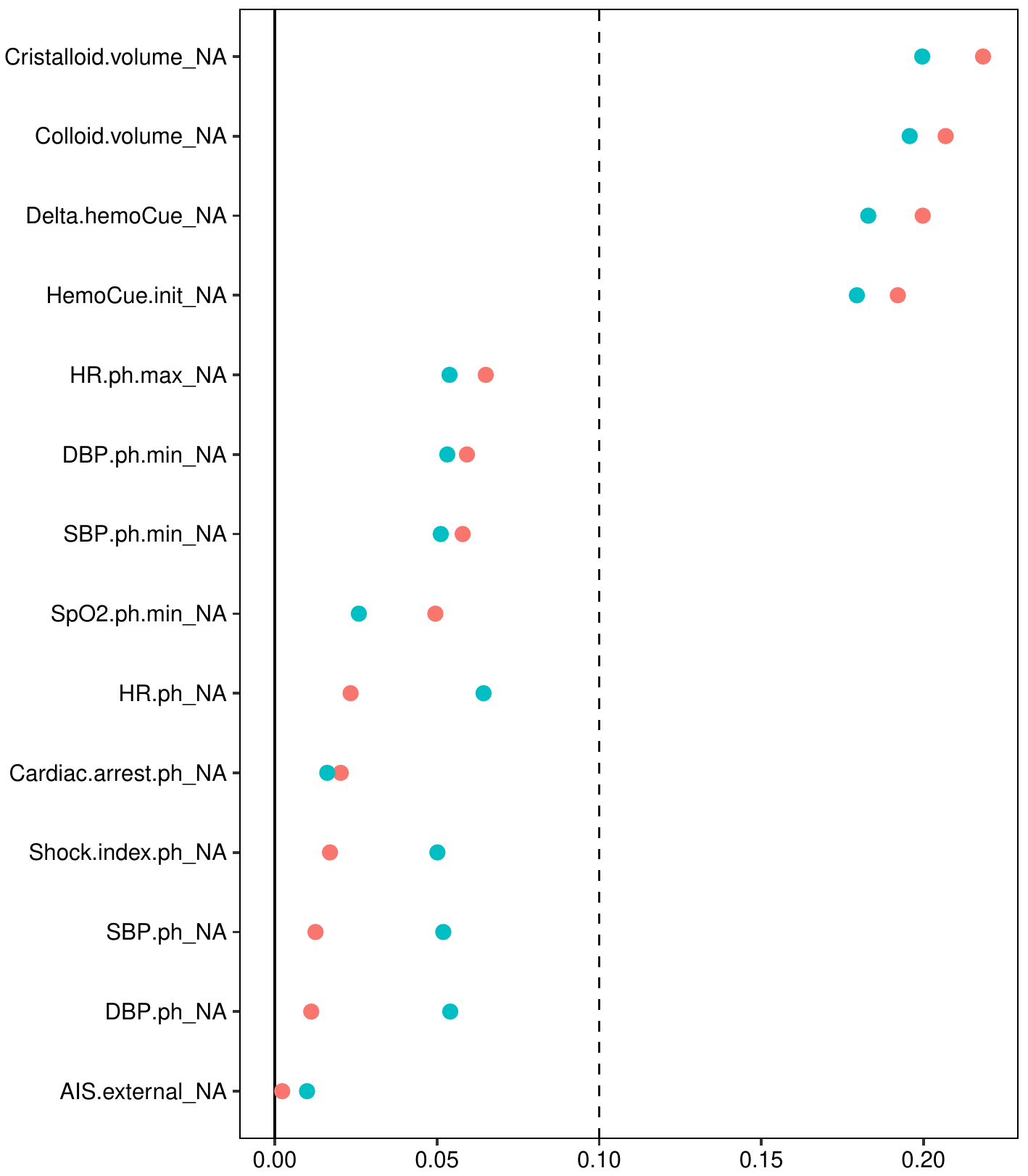}
         \caption{MICE}
         \label{fig:mice-balanceNA}
     \end{subfigure}
        \caption{Absolute difference in proportion for observed and missing values. Red: before adjustment, blue: after adjustment.}
        \label{fig:balanceNA}
\end{figure}

We notice a large difference between the IPW and the AIPW estimations. The AIPW estimations seem more reasonable for two reasons: first, the medical experts have noticed beneficial effects of TA for some of their TBI patients in practice and a previous clinical trial, focussing on a slightly different patient group, has also exhibited a potential benefit from the drug for patients with TBI; moreover, the results of the clinical trial studying the effect of the drug on all TBI patients indicate that on average there is neither benefit nor harm in prescribing the drug \citep{cap2019crash}; second, for the AIPW estimators, we incorporate much more available information, namely all identified features that are strongly related to the outcome $Y$ according to the expert panel (blue nodes on Figure \ref{fig:dagitty}).
Finally, the compared estimates have similar standard errors and asymptotic confidence intervals which are also close to the estimated bootstrap confidence intervals (the latter are not reported in Figure \ref{fig:dr-traumabase}).

\section{Discussion and perspectives}
\label{sec:discussion}
\subsection{Two families of treatment effect estimators handling missing attributes}
We have stressed the dyadic classification of previously exposed methods that allow treatment effect estimation with missing attributes, both in theory and in practice. The class of methods that relies on assumptions about the missingness mechanisms for treatment effect identifiability is currently often used, in combination with IPW-type estimators. We have also proposed an AIPW formulation for the most popular method from the first class, namely multiple imputation. However, methods of this first class have limited applicability in practice, most importantly they exclude informative missing data; this is a drawback of all developed methods in this class.
The second class, relying on the generalized propensity score and a different unconfoundedness assumption, can handle arbitrary missingness mechanisms, in particular the case where MAR does not hold, but to the best of our knowledge, implementable and versatile methods in this class have not been proposed so far. 

In practice, if one can exclude smooth regression functions for the treatment assignment and the outcome model, such as logistic and linear models, and if the ``unconfoundedness despite missingness" assumption is likely to hold---for more details on this, we refer to \citet{blake_etal_2019}---we advocate our tree-based estimator $\hat{\tau}_{MIA}$ in its AIPW-form and its mean-imputation variant. If one is willing to make stronger (parametric) assumptions about the structure of $X$ and its relationship with $W$ and $Y$, then our second estimator $\hat{\tau}_{EM}$ can also be considered as an alternative.

\subsection{Heterogeneous treatment effects and policy learning} Instead of estimating the average treatment effect $\tau$, one could be interested in the conditional average treatment effect function, defined as $\tau(x) = \mathbb{E}[Y(1) - Y(0) \cond X=x]$, for several reasons. For instance one might be interested in estimating how treatment effects vary across sub-populations, or assessing whether there is heterogeneity in the population w.r.t. a given treatment. Such questions anticipate problems of learning decision rules that exploit treatment effect heterogeneity \citep{athey2017efficient}.
%
%
%

In light of our medical application, heterogeneous treatment effect estimation is of particular interest because of the known existing heterogeneity among traumatic brain injury patients in terms of clinical presentation, pathophysiology and outcome. It is even more relevant since to this date there is no general classification of patients with traumatic brain injury. Hence a causal inference approach allowing classification w.r.t. treatment heterogeneity for any given treatment is of interest for practitioners in critical care management. 

\subsection{Weighted Treatment Effects}

Throughout this paper, we have focused on cases with overlap \eqref{eq:overlap}, i.e., where all units have
a realistic chance of being randomized to both treatment and control. In some cases, however, there
may be subjects who are quasi-deterministically assigned to one of the two treatment arms---in which case
the methods developed here may be unstable and/or have very large variance. When this happens, it is
common to shift focus away from the average treatment effect, and towards alternative weighted estimands
that are more robust to lack of overlap. For example, if some units are quasi-deterministically assigned to control
(but no units are quasi-deterministically assigned to treatment, i.e., propensity scores are uniformly bounded below 1),
then estimating the average treatment effect on the treatment is a popular way to avoid overlap problems \citep{imbens_2004}.
\citet{crump_etal_2009} and \citet{li_etal_JASA2018} discuss other weighted estimands that can be used
when overlap problems get more severe and propensity scores may get arbitrarily close to both 0 and 1.

Although we do not discuss it here, the arguments developed in this paper can be applied directly to estimators of other
weighted treatment effects. We implement extensions of the random forest based estimator described in \ref{sec:nonparam}
for estimating both the average treatment effect and the overlap-weighted treatment effect of \citet{li_etal_JASA2018}
in the \texttt{R} package \texttt{grf} \citep{grf}.

\subsection{Further identification strategies}

Although the two lines of approaches studied here for identification of average treatment effects with missing
attributes are the most prevalent in applied work, they are far from exhaustive.
For example, \citet{yang_etal_2019} consider a setting with outcome-independent missingness, $Y_i \indep R_i \cond \cb{X_i, \, W_i}$,
and find that $\tau$ can be identified via a set of integral equations.
We expect the area of methods development for causal inference with missing attributes to be a fruitful
research area for years to come.

\paragraph{Acknowledgement} We thank Jean-Pierre Nadal for fruitful discussion, Helen Blake and Julie Tibshirani for their suggestions for the simulation study, and the Delphi expert committee for the medical insight and advice on traumatic brain injury and hemorrhagic shock. We acknowledge funding from the EHESS PhD fellowship.


\begin{supplement}[id=suppA]
\sname{Supplementary material}
\stitle{Further simulation results and details on the Traumabase}
\slink[url]{https://bit.ly/3fWu9g0}
\sdescription{In this material we show additional simulation results, including different simulation scenarios and estimators. Furthermore we provide a glossary for the Traumabase variables and an additional analysis on this data set.}
\end{supplement}

\bibliographystyle{imsart-nameyear}
\bibliography{causal_inference_biblio}

\end{document}